\documentclass[pra, aps, twocolumn, groupedaddress, superscriptaddress]{revtex4}
\usepackage{slashed}
\usepackage{graphicx}
\usepackage{subfigure}
\usepackage[usenames, dvipsnames]{color}
\usepackage{graphics,amsfonts,amsmath, amssymb}
\usepackage{hyperref}
\usepackage{bm}
\usepackage{amsfonts}
\usepackage{lipsum}

\hypersetup{backref,
colorlinks=true,
urlcolor=blue,
linkcolor=blue,
linktoc=page,
citecolor=blue}

\everymath{\displaystyle} 
\begin{document}

\title{Universal evolution of non-classical correlations due to collective spontaneous emission}

\author{A. Slaoui}
\affiliation{LPHE-Modeling and Simulation, Faculty of Sciences, University Mohammed V, Rabat, Morocco.}
\email{abdallahsalaoui1992@gmail.com}
\author{M. I. Shaukat}
\affiliation{CeFEMA, Instituto Superior T\'ecnico, Universidade de Lisboa, Lisboa, Portugal}
\affiliation{University of Engineering and Technology, Lahore (RCET Campus), Pakistan}
\email{muzzamalshaukat@gmail.com}
\author{M. Daoud}
\affiliation{Department of Physics , Faculty of Sciences-Ain Chock, University Hassan II, Casablanca, Morocco.}
\affiliation{Abdus Salam International Centre for Theoretical Physics, Miramare, Trieste, Italy.}
\email{m\_daoud@hotmail.com}
\author{R. Ahl Laamara}
\affiliation{LPHE-Modeling and Simulation, Faculty of Sciences, University Mohammed V, Rabat, Morocco.}
\affiliation{Centre of Physics and Mathematics, CPM, CNESTEN, Rabat, Morocco.}
\email{ahllaamara@gmail.com}


\begin{abstract}
	We explore the spontaneous generation and decay of quantum correlations between two identical atoms coupled to a common Markovian environment in the presence of electromagnetic field modes. For this purpose, we analyze the dynamics of quantum correlations by employing the concurrence, the trace quantum discord and the local quantum uncertainty, for collective Dicke states. It is shown that the collective damping and dipole-dipole interaction plays a key role in enhancing non-classical correlations during the process of intrinsic decoherence. The quantum correlations can be maintained over a long time but for small distance between the two atoms.\par
\end{abstract}

\maketitle

\section{Introduction}
Quantum information theory (QIT) is an exciting field which lies at the intersection of physics, mathematics and computer science. It deals with the storage, transmission and processing of information using quantum-mechanical systems. Entanglement  is  the  key  resource of many QIT applications \cite{Einstein1935,Bell1964,Hill1997} like quantum  key  distribution \cite{Ekert1991}, quantum  teleportation \cite{Bennett1992} and  quantum  dense coding  \cite{Bennett1993}. \par
In this sense, quantification of quantum correlations in multipartite systems has attracted a lot of interest in the literature. Various quantum correlations quantifies were introduced. The first one is quantum discord (QD) \cite{Olliver2001} which has been shown a fundamental resource in quantum information processing \cite{Modi2012}. Despite of an important effort concerning the comparison of entanglement and QD for several families of quantum states \cite{Luo2008, Ali2010, Tanaś2013}, scientists, unfortunately, have not yet described the clear evidence of relation between them \cite{Céleri2011}. Recent studies showed that quantum discord act as a resource of entanglement distribution \cite{Chuan2012, Streltsov2012} and a quantitative measure in quantum-state merging \cite{Madhok2011, Cavalcanti2011}.
\par
One of the major challenges for the physical realization of quantum information and computation protocols is the decoherence which arises due to the quantum system coupling with its surroundings, causing loss of information from the system to its environment. During 20th century, it has been perceived that the spontaneous emission of multi-atomic system can be altered due to the collective properties of the system, in comparison to single atom case. Two different spontaneous decay rates (superradiant and subradiant) due to inter-atomic dipole interaction has been recognized by Dicke \cite{Dicke1954}. Tavis
and Cummings \cite{Tavis1968} studied the interaction between the atoms and a single-mode radiation field inside a cavity, as a partuclar case of Dicke Model.
\par
There are many physical systems suggested for the implementation of QIT such as cavity QED \cite{Lukin2001}, ion traps \cite{Steane1997} and quantum dots \cite{Peter2017}. The time evolution of entanglement for a system of two qubits, or two-level atoms, has been widely studied in recent years \cite{Braun2002, Yu2004, Ficek2006, Muzzamal2013, Ficek2008, Verstraete2009}. The investigation on the collective Dicke states has been carried out to measure qubit-qubit entanglement \cite{muzzamal2018, He2017}. Very recently, a long distance entanglement has been generated by Muzzamal et al., \cite{Muzzamal2018} by using quantum dark-soliton qubit in quasi one dimensional Bose-Einstein condensates \cite{Muzzamal2017}. The double or zero excitation bell states (pure or mixed) has been proposed in Ref. \cite{Khulud2011}
 
\par
More recently, it has been shown that the perfect communication of measurement results using classical tools is impossible when the measurement device is prepared in a classical state. In this process, the absence of quantum correlation between the measurement apparatus and the system of interest induces a lost of information. In addition, the study of non-classical correlations in multipartite systems continues to be an important issue in the literature \cite{Wang2010, Maziero2009, Bellomo2012, Pinto2013, Aaronson2013, Hu2014}.. It has been shown that quantum discord is a special kind of quantum correlation which gives beyond entanglement. Furthermore, this type of non-classical correlation is a valuable resource for several quantum protocols as for instance in quantum computation \cite{Knill1998}.


Another interesting family of states is a family of the two-qubit X-states, which are of interest here. Ali et al. \cite{Ali2010} used this state to make a closed form solution for quantum discord. However, it turned out that their algorithm is not universal. Later, Lu et al. \cite{Lu2011} proved that it is not possible to find a universal set of orthogonal projective measurements for the full family of X states. Some counterexamples have been given in \cite{Lu2011, Chen2011}. Instead, Chen et al. \cite{Chen2011}, confirmed the applicability of the algorithm for several special cases of X states. Recently, it has been shown that different measures of quantum correlations behave differently in their evolution \citep{Tanaś2013}. 
\par
The present work is organized as follows: In sec.  II, we discuss the main quantifiers of non-classical correlations, with special attention to the concurrence, the trace quantum discord and the local quantum uncertainty. We obtain an explicit formula of local quantum uncertainty and give also the analytical expression of the geometric discord based on the trace norm in a two-qubit X state. The theoretical model consisting of two-two level atoms is discribed in sec. III, where we also present the Markovian master equation and extract the density matrix elements for the Dicke states to evaluate the quantum correlations quantifiers (concurrence, trace quantum discord and the local quantum uncertainty). Finally, we present the conclusion of the present investigation in Sec. IV

\section{Quantifiers of quantum correlations}
Entanglement is a useful physical resource of QIT and describes the correlations between quantum systems that is much stronger and richer than any classical correlation. The study of entanglement and in particular how it can be quantified is a central topic within QIT. Therefore, the quantitative measures of the entanglement in bipartite and multipartite quantum systems are the entanglement of formation \cite{Bennett1996,Popescu1997}, concurrence \cite{Wootters2001,Yu2009}, linear entropy \cite{Bose2000}, entanglement of distillation \cite{Brassard1996}, and negativity \cite{Peres1996,Vidal2002}. But, a most widely accepted measure for a two qubit system is the concurrence defined by Wootters \cite{Wootters1998},
\begin{equation} 
C\left( \rho\right)  = \max \left\{ {0,\sqrt{\vartheta _1} - \sqrt{\vartheta _2} - \sqrt{\vartheta _3} - \sqrt{\vartheta _4}} \right\}, \label{C}
\end{equation}
with ${\vartheta _i}$'s are the eigenvalues (decreasing order) of the Hermition matrix $R=\rho \tilde \rho $, where the spin flip density matrix $\tilde \rho  = \left( {{\sigma _y} \otimes {\sigma _y}} \right){\rho ^*}\left( {{\sigma _y} \otimes {\sigma _y}} \right)$, with ${\rho ^*}$ and ${\sigma _y}$ being the complex conjugate of $\rho$ and the Pauli matrix, respectively.\\
An alternative approach to investigate the quantum correlation for an arbitraray state has been proposed by Zurek et al., \cite{Ollivier2001} and Vedral et al., \cite{Henderson2001}; It is called "quantum discord" and defined as the difference between two classically-equivalent expressions of the mutual information, that is to say the original quantum mutual information $\mathcal{I}\left( {{\rho_{AB}}} \right): = S\left( {{\rho_{A}}} \right) + S\left( {{\rho_{B}}} \right) - S\left( {{\rho_{AB}}} \right)$, and the local measurement-induced quantum mutual information $\mathcal{C}\left( {{\rho_{AB}}} \right)$;
\begin{equation}
{\cal Q}\left( {{\rho _{AB}}} \right)= {\cal I}\left( {{\rho _{AB}}} \right) - \mathop {\max }\limits_{{\pi _B}^j} \left( {S\left( {{\rho _B}} \right) - \sum\limits_j {{p_{B,j}}S\left( {{\rho _{^{B,j}}}} \right)} } \right),
\end{equation} 
where $S\left( \rho  \right) =  - tr\left( {\rho \log_{2}\rho } \right)$ is the von Neumann entropy, ${{\pi _B}^j}$ is a set of local projective measurements on the subsystem $B$, and ${{\rho _{^{B,j}}}}$ is the conditional state of system $B$ associated with outcome $j$.
\par
An analytical approach to evaluate the entropic discord is in general a difficult task  due to an optimization procedure for the conditional entropy over all local generalized measurements, even for the simplest case of two-qubit system. These difficulties led Dakic et al., to propose a geometric measure of quantum discord in terms of its minimal Hilbert-Schmidt norm (Schatten $p$-norms) distance from the set of classical states \cite{Dakic2010}. Despite its casiness of computability \cite{Bellomo1,Bellomo2,DaoudPLA,DaoudIJQI}, this measure is not a good measure of quantum correlations for $p\textgreater1$, since it may increase under local reversible operations on the unmeasured subsystem, and also it is non contractible under trace preserving channels \cite{piani}. The Bures norm (trace norm with $p= 1$) is the only Schatten $p$-norm which is contractible \cite{paula2013,Bromley2014}. Therefore, the trace distance quantum discord (TQD) for a two-qubit state $\rho$ is defined by:
\begin{eqnarray}\label{eq1}
D_{\rm T}(\rho)= \min_{\chi\in\Omega}||\rho-\chi||_1,
\end{eqnarray}
where $||\rho-\chi||_1={\rm Tr}\sqrt{(\rho-\chi)^\dag (\rho-\chi)}$, and the classical-quantum state $\chi = \sum_k p_k ~\Pi_{k,1}\otimes \rho_{k,2}$ belongs to the set $\Omega$ of classical-quantum states with $\Pi_{k,1}$ and $\rho_{k,2}$ denoting a set of orthogonal projectors for subsystem $1$ and  being a general density matrix associated with the second qubit, respectively. The minimization over the whole set of classical states for 1-norm two-qubit $X$ states has been proposed in Ref. \cite{Ciccarello2014}. Thus, the $X$-state density matrix is of form
\begin{equation}
\rho  = \left( {\begin{array}{*{20}{c}}
	{{\rho _{11}}}&0&0&{{\rho _{14}}}\\
	0&{{\rho _{22}}}&{{\rho _{23}}}&0\\
	0&{{\rho _{32}}}&{{\rho _{33}}}&0\\
	{{\rho _{41}}}&0&0&{{\rho _{44}}}
	\end{array}} \right). \label{X}
\end{equation}
The phase factors ${\rho _{14} \mathord{\left/
		{\vphantom {\rho _{14} {\left| \rho _{14}\right|}}} \right.
		\kern-\nulldelimiterspace} {\left| \rho _{14} \right|}} = e^{i\theta _{14}}$ and ${\rho _{23} \mathord{\left/
		{\vphantom {\rho _{23} {\left| \rho _{23}\right|}}} \right.
		\kern-\nulldelimiterspace} {\left| \rho _{23} \right|}} = e^{i\theta _{23}}$ of the off diagonal elements can be removed using the local unitary transformations acting on the two qubits of the system
	\begin{equation*}
	{\left| 0 \right\rangle _k} \to \exp \left( {\frac{i}{2}\left( {{\theta _{14}} + {{\left( { - 1} \right)}^k}{\theta _{23}}} \right)} \right){\left| 0 \right\rangle _k}, \hspace{0.5cm} k=1,2
	\end{equation*}
with the unchanged rank and positive off-diagonal entries of the density matrix $\rho$, i.e., 
\begin{equation}
\rho  \to \hat \rho  = \left( {\begin{array}{*{20}{c}}
	{{\rho _{11}}}&0&0&{\left| {{\rho _{14}}} \right|}\\
	0&{{\rho _{22}}}&{\left| {{\rho _{23}}} \right|}&0\\
	0&{\left| {{\rho _{23}}} \right|}&{{\rho _{33}}}&0\\
	{\left| {{\rho _{14}}} \right|}&0&0&{{\rho _{44}}}
	\end{array}} \right),  \label{X1}
\end{equation} 
In the Fano-Bloch representation, Eq. (\ref{X1}) can be written as 
\begin{equation}\label{Fano-Bloch}
\hat\rho  = \frac{1}{4}\sum\limits_{\alpha ,\beta } {\mathcal{R}_{\alpha \beta }} {\sigma _\alpha } \otimes {\sigma _\beta },
\end{equation}
where the non vanishing correlation matrix elements $\mathcal{R}_{\alpha \beta } = tr\rho \left( {{\sigma _\alpha } \otimes {\sigma _\beta }} \right)$ are given by
\begin{eqnarray}
{\mathcal{R}_{11}}&=&2(|\rho_{23}|+ |\rho_{14}|), \qquad 
{\mathcal{R}_{22}}=2(|\rho_{23}|- |\rho_{14}|), \nonumber \\
{\mathcal{R}_{33}}&=&1-2(\rho_{22}+\rho_{33}),   \qquad  
{\mathcal{R}_{30}}=2(\rho_{11}+\rho_{22})-1, \nonumber \\
\mathcal{R}_{03}&=&2(\rho_{11}+\rho_{33})-1.
\end{eqnarray}
According to Ref. \cite{Ciccarello2014}, the trace distance quantum discord is invariant under local transformations and takes the form
\begin{eqnarray}\label{eq3}
D_{\rm T}(\rho)=\sqrt{\frac{\mathcal{R}_{11}^2
		\mathcal{R}_{\rm max}^2-\mathcal{R}_{22}^2\mathcal{R}_{\rm min}^2}
	{\mathcal{R}_{\rm max}^2-\mathcal{R}_{\rm min}^2+\mathcal{R}_{11}^2-\mathcal{R}_{22}^2}}, \label{D}
\end{eqnarray}
where 
$\mathcal{R}_{\rm min}^2=\min\{\mathcal{R}_{11}^2,\mathcal{R}_{33}^2\} \hspace{0.2cm} {\rm and}\hspace{0.2cm} \mathcal{R}_{\rm max}^2={\rm \max}\{\mathcal{R}_{33}^2,\mathcal{R}_{22}^2+\mathcal{R}_{30}^2\}.$
\par 
Very recently, a discord-like measure of quantum correlation (local quantum uncertainty (LQU)) formalized by Girolami et al \cite{Girolami2013} is defined as the minimum skew information achievable with a single local measurement \cite{Wigner1963}. It constitutes an alternative tool to evaluate the analytical expressions of quantum correlations encompassed in any bipartite systems. This measurement satisfies all the known criteria for a discord-like quantifier and also deeply related to quantum Fisher information in the context of quantum metrology \cite{Luo2003}. The local quantum uncertainty is given by
\begin{equation}
\mathcal{U}(\rho) \equiv \min_{K_A} \mathcal{I}(\rho, K_A \otimes
\mathbb{I}_B), \label{LQU}
\end{equation}
where $K_A$ is some local observable on subsystem $A$, and $\mathcal{I}(\rho,   K_A \otimes \mathbb{I}_B)$ is the skew information of the density operator $\rho$, i.e.,
\begin{equation}
\mathcal{I}(\rho,   K_1 \otimes
\mathbb{I}_2)=-\frac{1}{2}{\rm
	Tr}([\sqrt{\rho}, K_1 \otimes
\mathbb{I}_2]^{2}).
\end{equation}
For the bipartite $2\otimes d$ systems, Girolami et al have derived a closed form of LQU \cite{Girolami2013}:
\begin{equation}
\mathcal{U}(\rho) = 1 - {\rm max}\{ \xi_1, \xi_2, \xi_3\},  \label{LQU1}
\end{equation}
where $\xi_{i}$'s are the eigenvalues of the $3\times3$ symmetric matrix $W$ whose matrix elements are defined by,
\begin{equation}\label{w-elements}
\omega_{ij} \equiv  {\rm
	Tr}\{\sqrt{\rho}(\sigma_{i}\otimes
\mathbb{I}_B)\sqrt{\rho}(\sigma_{j}\otimes \mathbb{I}_B)\}, 
\end{equation}
with $i,j = 1, 2, 3$. For the $X$-type states, the matrix elements  of Eq. (\ref{w-elements}) are given by (see appendix A):

\begin{eqnarray}
{w_{11}} &=& \left( {\sqrt {{\lambda _1}}  + \sqrt {{\lambda _4}} } \right)\left( {\sqrt {{\lambda _2}}  + \sqrt {{\lambda _3}} } \right)   \nonumber \\
&+&{\frac{{\left( {T_{11}^2 - T_{22}^2} \right) + \left( {T_{12}^2 - T_{21}^2} \right) + \left( {T_{03}^2 - T_{30}^2} \right)}}{{4\left( {\sqrt {{\lambda _1}}  + \sqrt {{\lambda _4}} } \right)\left( {\sqrt {{\lambda _2}}  + \sqrt {{\lambda _3}} } \right)}}}, 
\nonumber \\
{w_{22}} &=& \left( {\sqrt {{\lambda _1}}  + \sqrt {{\lambda _4}} } \right)\left( {\sqrt {{\lambda _2}}  + \sqrt {{\lambda _3}} } \right) \nonumber \\
&+&\frac{{\left( {T_{22}^2 - T_{11}^2} \right) + \left( {T_{21}^2 - T_{12}^2} \right) + \left( {T_{30}^2 - T_{03}^2} \right)}}{{4\left( {\sqrt {{\lambda _1}}  + \sqrt {{\lambda _4}} } \right)\left( {\sqrt {{\lambda _2}}  + \sqrt {{\lambda _3}} } \right)}}, \nonumber \\
{w_{33}} &=& \frac{1}{2}\left[ {{{\left( {\sqrt {{\lambda _1}}  + \sqrt {{\lambda _4}} } \right)}^2} + {{\left( {\sqrt {{\lambda _2}}  + \sqrt {{\lambda _3}} } \right)}^2}} \right]  \nonumber \\
&+&\frac{{{{\left( {{T_{30}} + {T_{03}}} \right)}^2} - {{\left( {{T_{11}} - {T_{22}}} \right)}^2} - {{\left( {{T_{12}} + {T_{21}}} \right)}^2}}}{{8{{\left( {\sqrt {{\lambda _1}}  + \sqrt {{\lambda _4}} } \right)}^2}}} \nonumber  \\
&+&\frac{{{{\left( {{T_{03}} - {T_{30}}} \right)}^2} - {{\left( {{T_{11}} + {T_{22}}} \right)}^2} - {{\left( {{T_{12}} - {T_{21}}} \right)}^2}}}{{8{{\left( {\sqrt {{\lambda _2}}  + \sqrt {{\lambda _3}} } \right)}^2}}}, \nonumber \\
{w_{12}} &=& {w_{21}}  = \frac{1}{2}\frac{{{T_{11}}{T_{21}} + {T_{22}}{T_{12}}}}{{\left( {\sqrt {{\lambda _1}}  + \sqrt {{\lambda _4}} } \right)\left( {\sqrt {{\lambda _2}}  + \sqrt {{\lambda _3}} } \right)}}, \nonumber \\
{w_{13}} &=& {w_{31}} = {w_{23}} = {w_{32}} = 0, \label{w11}
\end{eqnarray}
where $T_{\alpha \beta } = tr\rho \left( {{\sigma _\alpha } \otimes {\sigma _\beta }} \right)$, and $\lambda_i (i = 1, 2, 3, 4)$ are the eigenvalues of the density matrix $\rho$ of Eq. (\ref{X}). We have to note here that when all the elements of the density matrix $\rho$ are real, these correlation matrix elements $T_{\alpha \beta }$ are coincident with the correlation matrix elements of the density matrix $\hat\rho$ occurring in Eq. (\ref{Fano-Bloch}). To analyze the system in an entangled or separable states, the local quantum uncertainty might be compared to the  concurrence or TQD described by Eq.'s (\ref{C}) and (\ref{D}), respectively.
\section{Theoretical Model and Master equation}
In this work, the system under consideration consists of two identical atoms with non-overlapping states, located at positions $r_{i}(i=1,2)$ having ground state $\left| g_{i} \right\rangle$ and excited state $\left| e_{i} \right\rangle$. The atoms are connected by dipole transition moments $\vec \mu$ and are coupled to all modes of the quantized electromagnetic field \cite{Agarwal1974,Tanas2004,Ficek1987,Auyuanet2010}. In the interaction picture, the Hamiltonian of the system after employing rotating wave approximation can be written as
\begin{equation}
\hat H = \hbar {\omega _0}{S^z} + \sum\limits_{\vec ks} {{\omega _k}\hat a_{\vec ks}^\dag {{\hat a}_{\vec ks}}}  - i\hbar \sum\limits_{\vec ks} {\left[ {\vec \mu .{{\vec g}_{\vec ks}}{S^ + }{{\hat a}_{\vec ks}} - H.c} \right]},
\end{equation}
where $S_i^ +  = \left| {{e_i}} \right\rangle \left\langle {{g_i}} \right|$ and $S_i^ -  = \left| {{g_i}} \right\rangle \left\langle {{e_i}} \right|$ represents the dipole raising and lowering operators, $S_i^z = \left| {{e_i}} \right\rangle \left\langle {{e_i}} \right| - \left| {{g_i}} \right\rangle \left\langle {{g_i}} \right|$ is the energy operator of the $i$th atom, ${{{\hat a}_{\vec ks}}}$ and ${\hat a_{\vec ks}^\dag }$ are the annihilation and creation operators with wave vector $\vec{k}$, frequency $\omega_k$ and the index of polarization $s$, respectively. The term ${{\vec g}_{\vec ks}}\left( {{{\vec r}_i}} \right)$ is the coupling constant given by:
\begin{equation}
{{\vec g}_{\vec ks}}\left( {{{\vec r}_i}} \right) = {\left( {\frac{{{\omega _k}}}{{2{\varepsilon _0}\hbar V}}} \right)^{\frac{1}{2}}}{{\vec e}_{_{\vec ks}}}{e^{i\vec k.{{\vec r}_i}}},
\end{equation} 
with $\vec e_{ks}$ denotes  the electric field polarization vector, $V$ is the quantization volume and ${\vec r}_i$ is the position of the $i$th atom.
\par
The master equation describing the evolution of the atomic density operator in the Born-Markov approximations has been derived by Lehmberg \cite{Lehmberg1970}. This derivation is a generalisation of the Lindblad master equation to the case of non-identical atoms (in this case, we have different transition frequencies ${{\omega _i}}$) interacting with a squeezed vacuum field \cite{Belavkin1969,Agarwal1970}, i.e,
\begin{equation}\label{master-eq}
\begin{array}{l}
\frac{{\partial \rho \left( \tau  \right)}}{{\partial t}} =  - i{\omega _0}\sum\limits_{i = 1}^2 {\left[ {S_i^z,\rho } \right]}  - i{\Omega _{12}}\sum\limits_{i \ne j}^2 {\left[ {S_i^ + S_j^ - ,\rho } \right]}  - \\
\hspace{1.2cm}\frac{1}{2}\sum\limits_{i,j = 1}^2 {{\Gamma _{ij}}\left( {\rho S_i^ + S_j^ -  + S_i^ + S_j^ - \rho  - 2S_j^ - \rho S_i^ + } \right)} ,
\end{array}
\end{equation}
where ${{\Gamma _{ij}}}(=\Gamma) $ is the damping by spontaneous emission for $i = j$, equal to the Einstein $A$ coefficient for spontaneous emission which is induced by the direct coupling of the atom with the radiation field, while ${{\Gamma _{ij}}}(=\gamma\Gamma) $ for $i\neq j$ depicts the collective damping resulting from mutual exchange of photons. The term ${\Omega _{ij}}(=\Omega)$ represents the interaction between two-two level atoms, defined by: 
\begin{equation}
{\Gamma _{ij}} = \frac{3}{2}\Gamma \left[ {\begin{array}{*{20}{l}}
	{\left[ {1 - {{\left( {\vec \mu .{{\vec r}_{ij}}} \right)}^2}} \right]\frac{{\sin \left( {{\xi _{ij}}} \right)}}{{{\xi _{ij}}}} + }\\
	{\left[ {1 - 3{{\left( {\vec \mu .{{\vec r}_{ij}}} \right)}^2}} \right]\left[ {\frac{{\cos \left( {{\xi _{ij}}} \right)}}{{{\xi _{ij}}^2}} - \frac{{\sin \left( {{\xi _{ij}}} \right)}}{{{\xi _{ij}}^3}}} \right]}
	\end{array}} \right],
\end{equation}
and
\begin{equation}
{\Omega _{ij}} = \frac{3}{4}\Gamma \left[ \begin{array}{l}
- \left[ {1 - {{\left( {\vec \mu .{{\vec r}_{ij}}} \right)}^2}} \right]\frac{{\cos \left( {{\xi _{ij}}} \right)}}{{{\xi _{ij}}}} + \\
\left[ {1 - 3{{\left( {\vec \mu .{{\vec r}_{ij}}} \right)}^2}} \right]\left[ {\frac{{\sin \left( {{\xi _{ij}}} \right)}}{{{\xi _{ij}}^2}} + \frac{{\cos \left( {{\xi _{ij}}} \right)}}{{{\xi _{ij}}^3}}} \right]
\end{array} \right].
\end{equation}
where
${\xi _{ij}} = {k_0}{r_{ij}} = {{2\pi {r_{ij}}} \mathord{\left/
		{\vphantom {{2\pi {r_{ij}}} {{\lambda _0}}}} \right.
		\kern-\nulldelimiterspace} {{\lambda _0}}}$, with ${{\lambda _0}}$ is the resonant wavelength, and ${r_{ij}} = \left| {{r_j} - {r_i}} \right|$ is the distance between the atoms.
\par
\begin{figure}[t!]
	\includegraphics[scale=0.5]{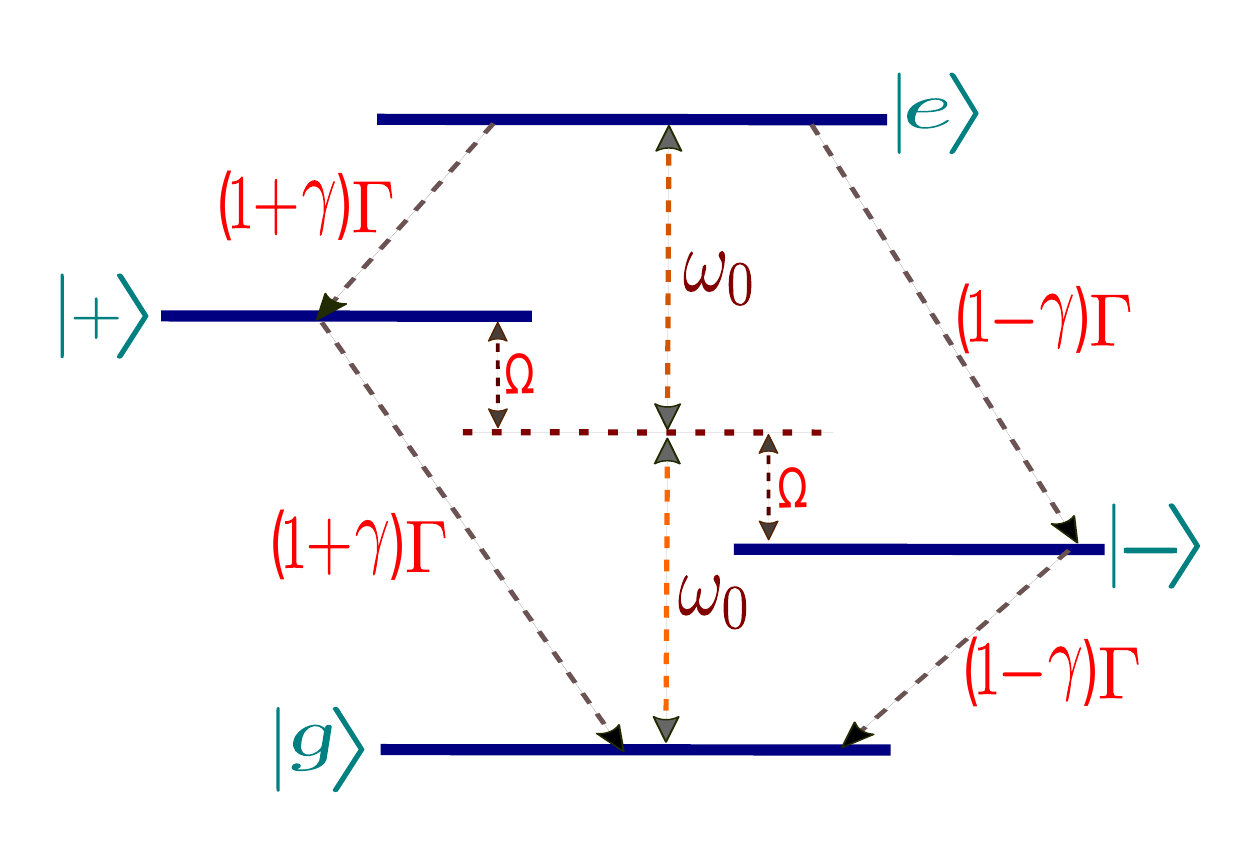}
	\caption{(color online) Energy-level diagram for the collective stats of two identical atoms, showing the frequency shifts $ \pm \hbar {\Omega _{12}}$ and decay constants $\left( {1 \pm \gamma } \right)\Gamma $.}
	\label{fig_scheme}
\end{figure}

The main concern of the present investigation is to study the evolution of the concurrence, the trace distance quantum discord and the local quantum uncertainty using the density-matrix formalism. To solve Eq. (\ref{master-eq}), we use the collective Dicke state representation, introduced by Dicke \cite{Dicke1954} where the two-atom system behaves as a single four-level system with states
\begin{eqnarray}
\left| g \right\rangle  &=& \left| {{g_1},{g_2}} \right\rangle, \nonumber \\
\left|  \pm  \right\rangle  &=& {{\left( {\left| {{e_1},{g_2}} \right\rangle  \pm \left| {{g_1},{e_2}} \right\rangle } \right)} \mathord{\left/
		{\vphantom {{\left( {\left| {{e_1},{g_2}} \right\rangle  \pm \left| {{g_1},{e_2}} \right\rangle } \right)} {\sqrt 2 }}} \right.
		\kern-\nulldelimiterspace} {\sqrt 2 }}, \nonumber \\
\left| e \right\rangle  &=& \left| {{e_1},{e_2}} \right\rangle, 
\end{eqnarray}
where, the energies correspond to respective states are ${E_g} =  - \hbar {w_0}$, ${E_ + } = \hbar {\Omega _{12}}$, ${E_ - } =  - \hbar {\Omega _{12}}$, and ${E_e} =  \hbar {w_0}$. Here, the states where $\left|  \pm  \right \rangle $ describes the maximally entangled symmetric and antisymmetric states, as schematically represented in Fig. (\ref{fig_scheme}).
With the use of Dicke or collective bases for arbitrary initial conditions, the elements of the density matrix $\rho$ can be determined by using master equation,
\begin{eqnarray}\label{rho-ee}
{\rho _{ee}}\left( \tau \right) &=& {e^{ - 2\tau }}{\rho _{ee}}\left( 0 \right), \nonumber \\
{\rho _{ +  + }}\left( \tau \right) &=& {e^{ - \left( {1 + \gamma } \right)\tau }}{\rho _{ +  + }}\left( 0 \right) \nonumber \\ 
&+& \frac{{\left( {1 + \gamma } \right)}}{{\left( {1 - \gamma } \right)}}\left( {{e^{ - \left( {1 + \gamma } \right)\tau }} - {e^{ - 2\tau }}} \right){\rho _{ee}}\left( 0 \right),    \nonumber \\
{\rho _{ -  - }}\left( \tau \right) &=& {e^{ - \left( {1 - \gamma } \right)\tau }}{\rho _{ -  - }}\left( 0 \right)
\nonumber \\ 
 &+& \frac{{\left( {1 - \gamma } \right)}}{{\left( {1 + \gamma } \right)}}\left( {{e^{ - \left( {1 - \gamma } \right)\tau }} - {e^{ - 2\tau }}} \right){\rho _{ee}}\left( 0 \right),  \nonumber \\
{\rho _{ +  - }}\left( \tau \right) &=& {e^{ - \left( {1 - 2i\eta } \right)\tau }}{\rho _{ +  - }}\left( 0 \right), \nonumber \\
{\rho _{eg}}\left( \tau \right) &=&{e^{ - \tau }}{\rho _{eg}}\left( 0 \right),
\end{eqnarray}
subject to the probability conservation ${\rho _{gg}} = 1 - {\rho _{ee}} - {\rho _{ +  + }} - {\rho _{ -  - }}$ with ${\rho _{jk}} = {\rho _{kj}}^ *$. Here, $\tau = \Gamma t$, $\gamma  = {{{\Gamma _{12}}} \mathord{\left/
		{\vphantom {{{\Gamma _{12}}} \Gamma }} \right.
		\kern-\nulldelimiterspace} \Gamma }$, and $\eta  = {{{\Omega _{12}}} \mathord{\left/
		{\vphantom {{{\Omega _{12}}} \Gamma }} \right.
		\kern-\nulldelimiterspace} \Gamma }$.
To measure the quantum correlations, we need the solutions for the density matrix elements in the standard product basis or into the Bell basis, for which we defined
$\left\{ {\left| 1 \right\rangle=\left| {{e_1},{e_2}} \right\rangle, \left| 2 \right\rangle=\left| {{e_1},{g_2}} \right\rangle, \left| 3 \right\rangle=\left| {{g_1},{e_2}} \right\rangle, \left| 4 \right\rangle=\left| {{g_1},{g_2}} \right\rangle} \right\}$ with
\begin{eqnarray}
\rho _{11}\left( \tau  \right) &=& {\rho _{ee}}\left( \tau  \right), \hspace{1cm}\rho _{14}\left( \tau  \right) = {\rho _{eg}}\left( \tau  \right), \nonumber \\
\rho _{22}\left( \tau  \right) &=& {{\left( {{\rho _{ +  + }}\left( \tau  \right) + {\rho _{ +  - }}\left( \tau  \right) + {\rho _{ -  + }}\left( \tau  \right) + {\rho _{ -  - }}\left( \tau  \right)} \right)} \mathord{\left/
		{\vphantom {{\left( {{\rho _{ +  + }}\left( \tau  \right) + {\rho _{ +  - }}\left( \tau  \right) + {\rho _{ -  + }}\left( \tau  \right) + {\rho _{ -  - }}\left( \tau  \right)} \right)} 2}} \right. 
		\kern-\nulldelimiterspace} 2}, \nonumber \\
\rho _{33}\left( \tau  \right) &=& {{\left( {{\rho _{ +  + }}\left( \tau  \right) - {\rho _{ +  - }}\left( \tau  \right) - {\rho _{ -  + }}\left( \tau  \right) + {\rho _{ -  - }}\left( \tau  \right)} \right)} \mathord{\left/
		{\vphantom {{\left( {{\rho _{ +  + }}\left( \tau  \right) - {\rho _{ +  - }}\left( \tau  \right) - {\rho _{ -  + }}\left( \tau  \right) + {\rho _{ -  - }}\left( \tau  \right)} \right)} 2}} \right.
		\kern-\nulldelimiterspace} 2},  \nonumber \\
\rho _{23}\left( \tau  \right) &=& {{\left( {{\rho _{ +  + }}\left( \tau  \right) + {\rho _{ +  - }}\left( \tau  \right) - {\rho _{ -  + }}\left( \tau  \right) - {\rho _{ -  - }}\left( \tau  \right)} \right)} \mathord{\left/
		{\vphantom {{\left( {{\rho _{ +  + }}\left( \tau  \right) + {\rho _{ +  - }}\left( \tau  \right) - {\rho _{ -  + }}\left( \tau  \right) - {\rho _{ -  - }}\left( \tau  \right)} \right)} 2}} \right.
		\kern-\nulldelimiterspace} 2}.  \nonumber \\	
\end{eqnarray}
Hereafter, we discuss the dependence of quantum correlations on different initial states. 
\subsection{Zero or Double Excitation}
Here, we discuss the decay of quantum correlation between two atoms , depending on the initially maximally entangled state of zero or double excitation.
\begin{equation}
	\left| {\psi \left( 0 \right)} \right\rangle  = \frac{1}{{\sqrt 2 }}\left( {\left| {{e_1},{e_2}} \right\rangle  + \left| {{g_1},{g_2}} \right\rangle } \right).
\end{equation}
Therefore, the density matrix becomes: 
\begin{equation}
\rho\left( \tau  \right) = \left( {\begin{array}{*{20}{c}}
	{a\left( \tau  \right)}&0&0&{d\left( \tau  \right)}\\
	0&{b\left( \tau  \right)}&e\left( \tau  \right)&0\\
	0&e\left( \tau  \right)&b\left( \tau  \right)&0\\
	{d\left( \tau  \right)}&0&0&{c\left( \tau  \right)}
	\end{array}} \right), \label{Matrixfirstcas}
\end{equation}
with
\begin{eqnarray}\label{a1cas}
a\left( \tau  \right) &=& \frac{1}{2}{e^{ - 2\tau }},  \hspace{2cm} b\left( \tau  \right) = \frac{e^{ - \tau }}{{2\left( 1 - \gamma ^2 \right)}}\delta,   \nonumber   \\
c\left( \tau  \right)&=& 1 - \frac{{{e^{ - 2\tau }}}}{2} - \frac{{{e^{ - \tau }}}}{{\left( {1 - {\gamma ^2}} \right)}}\delta,  \hspace{0.5cm}  d\left( \tau  \right) = \frac{1}{2}{e^{ - \tau }},  \nonumber   \\
e\left( \tau  \right) &=& \frac{{{e^{ - \tau }}}}{{2\left( {1 - {\gamma ^2}} \right)}}\left( {2\gamma Z 
 - \left( {1 + {\gamma ^2}} \right)\sinh \left( {\gamma \tau } \right)} \right),
\end{eqnarray}
where $Z={\cosh \left( {\gamma \tau } \right) - {e^{ - \tau }}} $ and $\delta= \left( 1 + \gamma ^2 \right)Z   - 2\gamma \sinh \left(\gamma \tau\right) $. We first consider the evaluation of the concurrence given by Eq. (\ref{C}). Depending on the largest eigenvalue, it's easy to verify that the concurrence is obtained as
\begin{equation}
C\left( \tau \right) = \left\{ \begin{array}{l}
\max \left\{ {0,{C_1}\left( \tau  \right)} \right\}{\rm{If}}\sqrt {{\vartheta _1}}  = \left| d\left( \tau  \right) + \sqrt{a\left( \tau  \right)c\left( \tau  \right)} \right|\\
\max \left\{ {0,{C_2}\left( \tau  \right)} \right\}{\rm{If}}\sqrt {{\vartheta _1}}  = \left| b\left( \tau  \right) + e\left( \tau  \right) \right|,\\
\max \left\{ {0,{C_3}\left( \tau  \right)} \right\}{\rm{If}}\sqrt {{\vartheta _1}}  = \left| b\left( \tau  \right) - e\left( \tau  \right) \right|, \label{con for des}
\end{array} \right.
\end{equation}
with
\begin{align}
{C_1}\left( \tau  \right) &= \frac{{{e^{ - \tau }}}}{{\left( {1 - {\gamma ^2}} \right)}}\left( {\left( {1 - {\gamma ^2}} \right) -\delta} \right), \nonumber \\
{C_2}\left( \tau  \right) &= \frac{{{e^{ - \tau }}}}{{\left( {1 - {\gamma ^2}} \right)}}\left( {2\gamma Z - \left( {1 + {\gamma ^2}} \right)\sinh \left( {\gamma \tau } \right)} \right) \notag \\&- \frac{{{e^{ - \tau }}}}{{\sqrt {1 - {\gamma ^2}} }}{\left[ {2\left( {1 - {\gamma ^2}} \right) - {e^{ - \tau }}\left( { 2\delta + \left( {1 - {\gamma ^2}} \right){e^{ - \tau }}} \right)} \right]^{\frac{1}{2}}}, \nonumber \\
{C_3}\left( \tau  \right) &= \frac{{{e^{ - \tau }}}}{{\left( {1 - {\gamma ^2}} \right)}}\left( {\left( {1 + {\gamma ^2}} \right)\sinh \left( {\gamma \tau } \right) - 2\gamma Z} \right) \notag\\&- \frac{{{e^{ - \tau }}}}{{\sqrt {1 - {\gamma ^2}} }}{\left[ {2\left( {1 - {\gamma ^2}} \right) - {e^{ - \tau }}\left( { 2\delta + \left( {1 - {\gamma ^2}} \right){e^{ - \tau }}} \right)} \right]^{\frac{1}{2}}}.
\end{align}
Subsequently, we determined the analytic evolution of the quantum discord based on the trace norm and the local quantum uncertainty. The non vanishing matrix correlations of Eq. (\ref{Matrixfirstcas}), in the Fano-Bloch representation, are given by
\begin{eqnarray}
T_{11} &=& \frac{{{e^{ - \tau }}}}{{\left( {1 - {\gamma ^2}} \right)}}\left( {\left( {1 - {\gamma ^2}} \right) + 2\gamma Z - \left( {1 + {\gamma ^2}} \right)\sinh \left( {\gamma \tau } \right)} \right), \nonumber \\
T_{22} &=& \frac{{{e^{ - \tau }}}}{{\left( {1 - {\gamma ^2}} \right)}}\left(  2\gamma Z - \left( {1 + {\gamma ^2}} \right)\sinh \left( {\gamma \tau } \right)-\left( {1 - {\gamma ^2}} \right) \right), \nonumber \\
{T_{33}} &=& 1-\frac{{2{e^{ - \tau }}}}{{\left( {1 - {\gamma ^2}} \right)}}\delta, \hspace{0.8cm} T_{30} = e^{- 2\tau } - 1 + \frac{e^{ - \tau }}{{\left( {1 - {\gamma ^2}} \right)}}\delta, \nonumber \\
\end{eqnarray}
where, $T_{03} = T_{30}$. It is simple to check that the difference $T_{22}^2+T_{30}^2-T_{33}^2$ remains positive always and $T_{\rm max}=T_{22}^2+T_{30}^2$, irrespective of the parameters ($\gamma$ and $\tau$) values. Using Eq. (\ref{D}), one has to treat separately the two cases. For ${T_{33}}^2 \geq {T_{11}}^2$, the trace distance discord can be simply written as
\begin{equation}
D_T\left(  \tau  \right) =\left| T_{11} \right|.
\end{equation}
For ${T_{33}}^2 \leq {T_{11}}^2$, the trace distance discord is given by
\begin{equation}
D_T\left( \tau  \right) = \sqrt {\frac{{T_{22}^2\left( {T_{11}^2 - T_{33}^2} \right) + T_{11}^2T_{30}^2}}{{T_{22}^2 - T_{33}^2 + T_{11}^2}}}.
\end{equation}
To determine the analytic expression of the  local quantum uncertainty from Eq. (\ref{LQU}), it is necessary to calculate the elements given by Eq. (\ref{w-elements}). After some simplifications, we obtain
	\begin{align}
w_{11} &= \sqrt { (\beta+X) (\beta+Y) } \nonumber \\ &+ \frac{{{e^{ - 2\tau }}\left( {2\gamma  Z  - \left( {1 + {\gamma ^2}} \right)\sinh \left( {\gamma \tau } \right)} \right)}}{{\left( {1 - {\gamma ^2}} \right)\sqrt {(\beta+X) (\beta+Y)} }}, \nonumber \\
w_{22} &= \sqrt { (\beta+X) (\beta+Y)} \nonumber \\ &- \frac{{{e^{ - 2\tau }}\left( {2\gamma Z  - \left( {1 + {\gamma ^2}} \right)\sinh \left( {\gamma \tau } \right)} \right)}}{{\left( {1 - {\gamma ^2}} \right)\sqrt {(\beta+X) (\beta+Y)} }},  \nonumber \\
w_{33} &= \frac{1}{2}\left( X+Y \right)+ \frac{{{{\left( {{e^{ - 2\tau }} - 1 + \beta} \right)}^2} - {e^{ - 2\tau }}}}{{2\left(  \beta +X  \right)}} \nonumber \\ &- \frac{{{e^{ - 2\tau }}{{\left( {2\gamma Z - \left( {1 + {\gamma ^2}} \right)\sinh \left( {\gamma \tau } \right)} \right)}^2}}}{{2{{\left( {1 - {\gamma ^2}} \right)}^2}\left( {\beta + Y
 } \right)}},
	\end{align}
with $X={1  + {e^{ - \tau }}\sqrt {1 - {e^{ - 2\tau }} - 2\beta} } $, $Y= { {e^{ - \tau }}\sqrt {1 + {e^{ - 2\tau }} - 2{e^{ - \tau }}\cosh \left( {\gamma \tau } \right)} }$ and
$\beta=e^{ - \tau }\delta/({1 - {\gamma ^2}}) $.
	\begin{figure}[h!]
		\centering
		\begin{minipage}[t]{3in}
			\centering
			\includegraphics[scale=0.55]{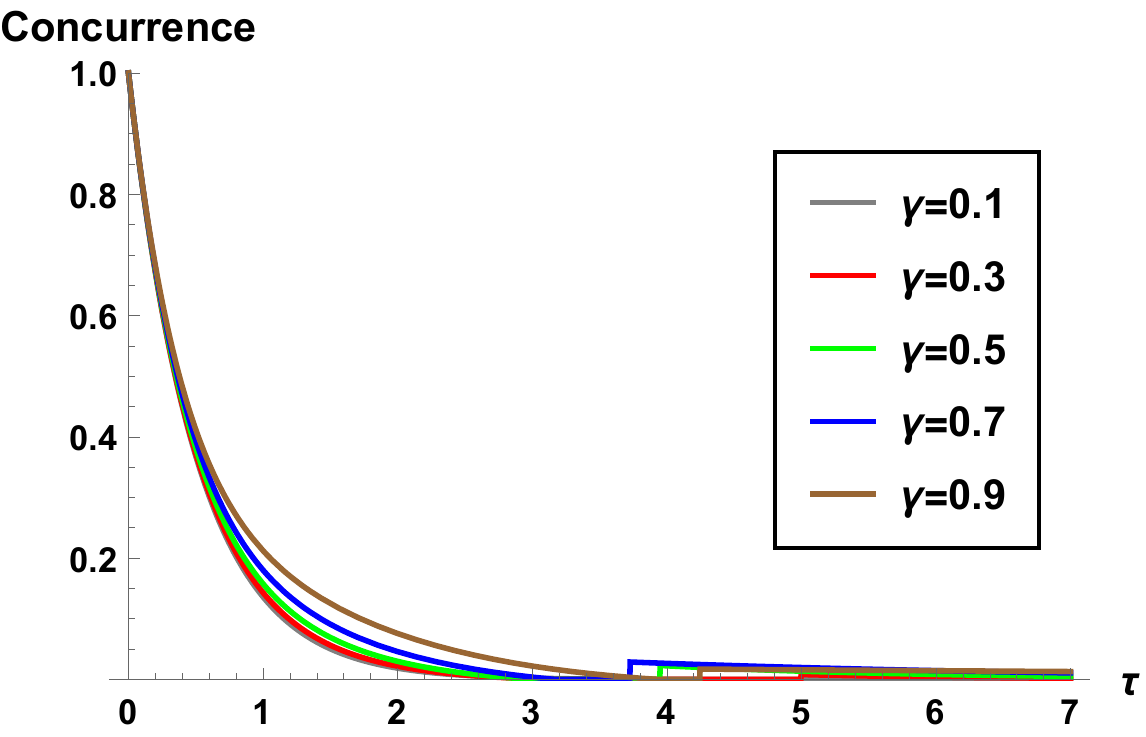}
		\end{minipage}
		\begin{minipage}[t]{3in}
			\centering
			\includegraphics[scale=0.55]{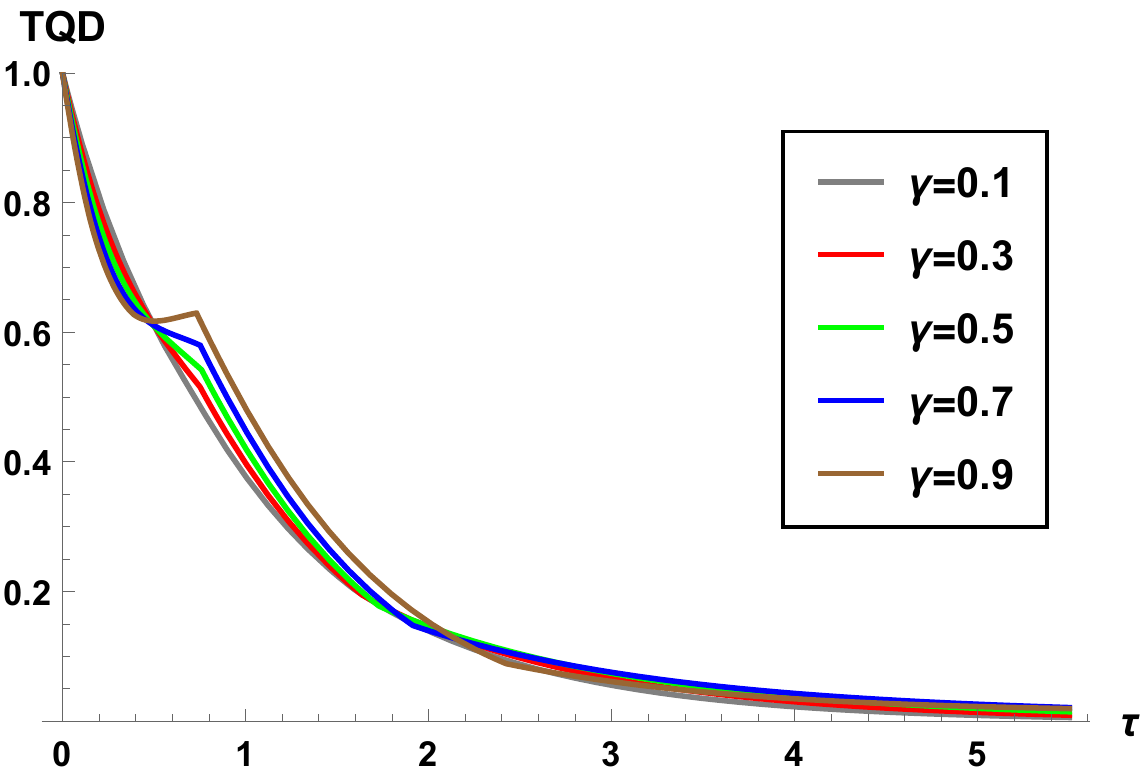}
		\end{minipage}
		\begin{minipage}[t]{3in}
			\centering
			\includegraphics[scale=0.52]{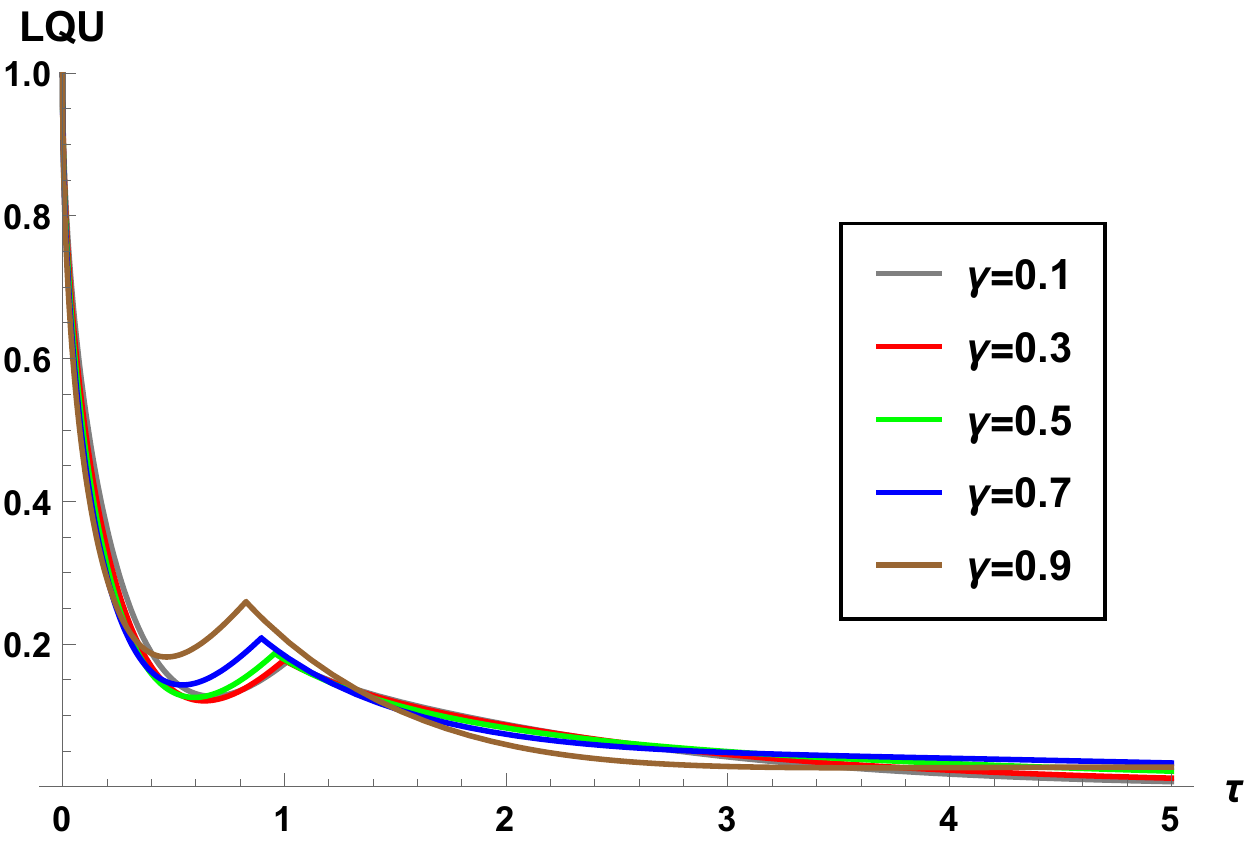}
		\end{minipage}\\
		\caption{Behavior of concurrence, trace quantum discord and local quantum uncertainty for two atoms maximally initially entangled versus the parameter $\tau$ for different values of $\gamma$.}
		\label{fig_Both excited states}
	\end{figure}
The dynamical evolution of quantum correlations when both the atoms are, initially, in an entangled state of zero or double excitation, is depicted in Fig. (\ref{fig_Both excited states}). It is observed that, the different measures of the quantum correlations, decays differently, where the concurrence, initialy, goes to zero (sudden death) and then revives over a long time with the dark period $ \Delta \tau $ which depends on each spontaneous emission parameter $\gamma$, but the other measures (TQD and LQU) follow the exponential decay of the correlation. Despite of the fact that concurrence is a good measure of correlation for two-level system, it can not capture all the nonclassical correlations in the system  with which other measures exponential decay make this system more worthwhile.
\subsection{Single-Atom Excitation.}
In this subsection, we describe the spontaneous generation of correlation between two atoms by assuming, initially, the separable state of singly excited state,  i.e., $\left| {{e_1},{g_2}} \right\rangle$. Thus, the density matrix elements are given by
\begin{eqnarray}
\rho _{22}\left( \tau  \right) &=& \frac{e^{ - \tau }}{2}\left( \cosh \left(\gamma \tau \right) + \cos \left( 2\eta \tau \right) \right), \nonumber  \\
\rho _{33}\left( \tau  \right) &=& \frac{e^{ - \tau }}{2}\left( \cosh \left(\gamma \tau \right) - \cos \left( 2\eta \tau \right) \right),  \nonumber  \\
\rho _{23}\left( \tau  \right) &=& \frac{e^{ - \tau }}{2}\left( i\sin \left(2\eta \tau \right) - \sinh \left(\gamma \tau \right) \right), \nonumber  \\
\rho _{44}\left( \tau  \right) &=& 1 - e^{ - \tau }\cosh \left(\gamma \tau \right).
 \label{eq64}
\end{eqnarray}
The nonvanishing correlation matrix elements are given by
\begin{eqnarray}
\mathcal{R}_{11} &=& \mathcal{R}_{22} = e^{ - \tau }\sqrt {\sin {\left(2\eta \tau \right)^2} + \sinh {\left(\gamma \tau\right)^2}},   \nonumber  \\
T_{11} &=& T_{22} = - e^{ - \tau }\sinh \left(\gamma\tau \right), \nonumber  \\
\mathcal{R}_{33} &=&T_{33} = 1 - 2e^{ - \tau }\cosh \left(\gamma \tau\right), \nonumber  \\
\mathcal{R}_{30} &=& T_{30} = e^{ - \tau }\left( \cosh \left(\gamma \tau \right) + \cos \left(2\eta \tau \right) \right) - 1, \nonumber  \\
\mathcal{R}_{03} &=& T_{03} = e^{ - \tau}\left( \cosh \left(\gamma \tau\right) - \cos \left( 2\eta \tau \right) \right) - 1.    \nonumber  \\
\end{eqnarray}
Based on the above formalism, the analytical expression to determine the concurrence and quantum discord is given by
\begin{equation}
C\left( \tau  \right) =D_{T}\left(\tau \right)= e^{-\tau }\sqrt{\sinh{\left(\gamma\tau \right)^2} + \sin\left(2\eta \tau\right)^2},
\end{equation}
whereas, the local quantum uncertainty is described by Eq. (\ref{w-elements}) with
\begin{eqnarray}
w_{11} &=& \frac{{{e^{ - \tau }}\left( {\cosh \left( {\gamma \tau } \right) + \cos \left( {2\eta \tau } \right)} \right)\left( {1 - {e^{ - \tau }}\cosh \left( {\gamma \tau } \right)} \right)}}{{\sqrt {{e^{ - \tau }}\cosh \left( {\gamma \tau } \right)\left( {1 - {e^{ - \tau }}\cosh \left( {\gamma \tau } \right)} \right)} }}, \nonumber \\
w_{22} &=& \frac{{{e^{ - \tau }}\left( {\cosh \left( {\gamma \tau } \right) - \cos \left( {2\eta \tau } \right)} \right)\left( {1 - {e^{ - \tau }}\cosh \left( {\gamma \tau } \right)} \right)}}{{\sqrt {{e^{ - \tau }}\cosh \left( {\gamma \tau } \right)\left( {1 - {e^{ - \tau }}\cosh \left( {\gamma \tau } \right)} \right)} }},  \nonumber \\
w_{33} &=& \frac{{2\cosh \left( {\gamma \tau } \right) - {e^{ - \tau }}\left( {1 + 2\sinh^2 {\left(\gamma \tau \right)} - \cos \left( {4\eta \tau } \right)} \right)}}{{2\cosh \left( {\gamma \tau } \right)}}, \nonumber \\
\end{eqnarray}
\begin{figure}[t!]
	\centering  
	\begin{minipage}[t]{3in}
		\centering
		\includegraphics[scale=0.55]{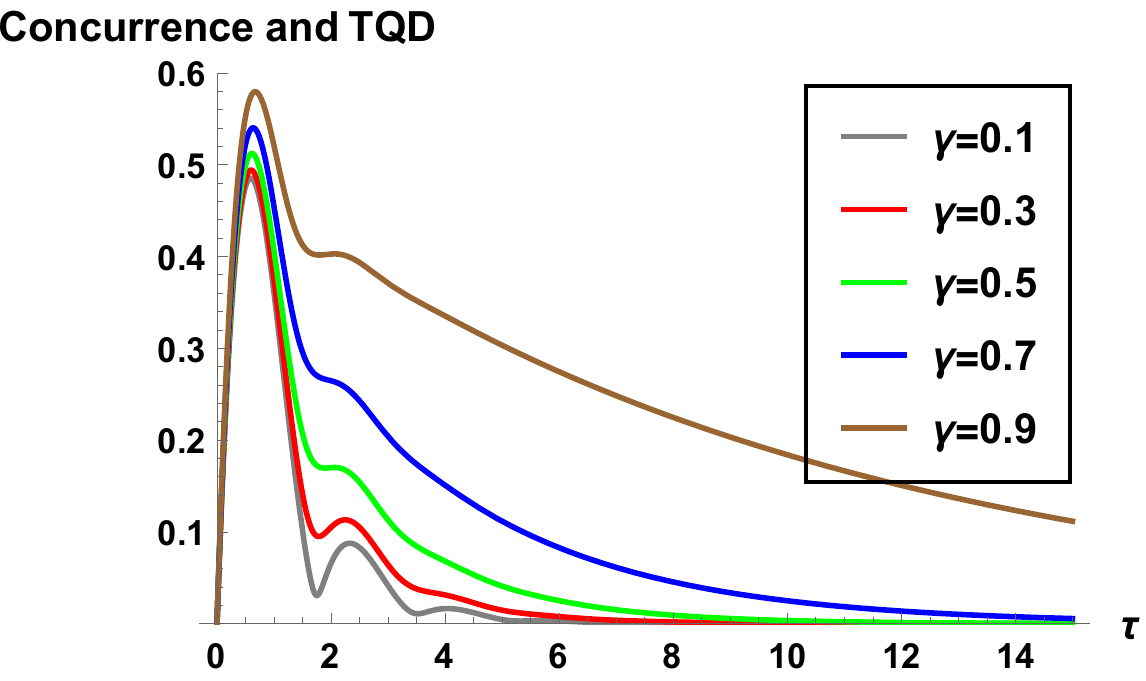}
	\end{minipage}\\
	\caption{Evolutions of concurrence and geometric quantum discord for Single-atom initial excitation versus the parameter $\tau$ for the different values of $\gamma$ with $\eta =0.9$.}\label{fig_1} 
\end{figure}
\begin{figure}[t!]
	\centering
	\begin{minipage}[t]{3in}
		\centering
		\includegraphics[scale=0.55]{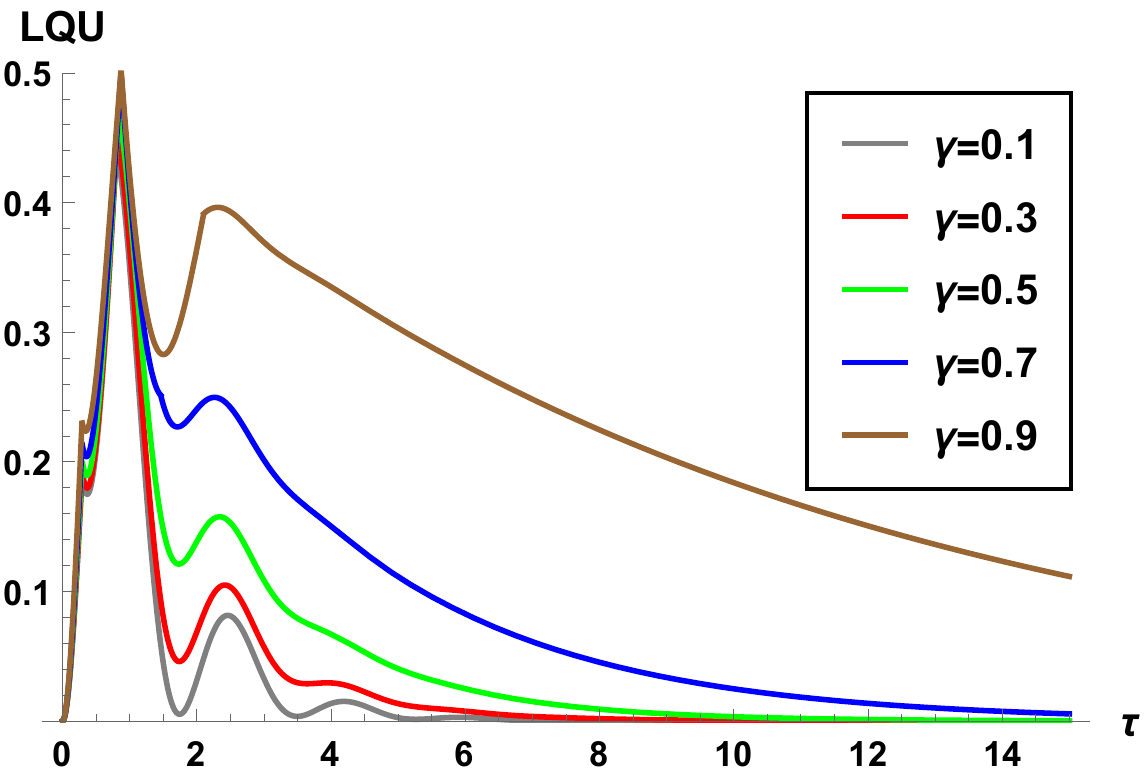}
	\end{minipage}\\
	\caption{Evolution of local quantum uncertainty for Single-atom initial excitation versus the parameter $\tau$ for the different values of $\gamma$ with $\eta =0.9$.}
\label{fig_4} 
\begin{minipage}[t]{3in}
		\centering
		\includegraphics[scale=0.55]{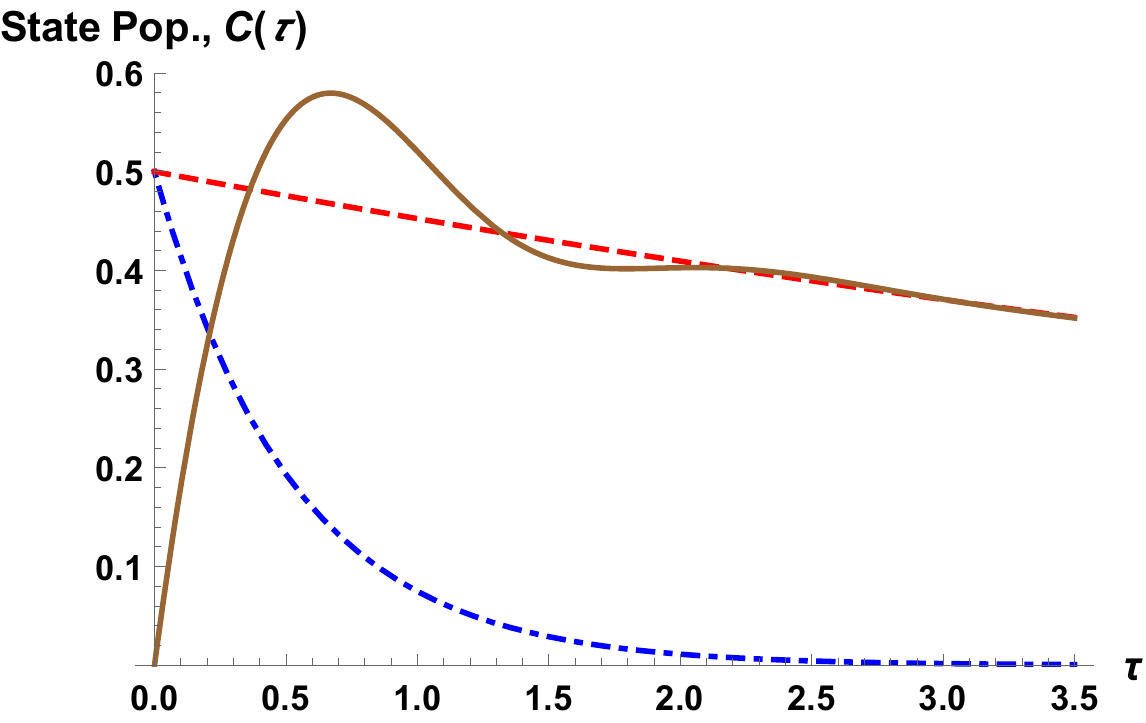}
	\end{minipage}\\
	\caption{Population of symmetric state $|+\rangle$ (dotted-dashed curve), antisymmetric state $|-\rangle$  (dashed curve) and time evolution of concurrence $C(\tau)$ (solid surve) at $\gamma =0.9$. }\label{fig_2} 
\end{figure}
Fig.'s (\ref{fig_1})-(\ref{fig_4}) display the spontaneous generation of quantum correlation by assuming, initially, the superposition of maximally entangled symmetric and anti-symmetric state. Contrary to the later case of initial entanglement, all the correlations depend on both the collective damping and dipole-dipole interaction term with which, firstly, displays a fast increase, being followed by a very slow oscillatory decay. Moreover, one can clearly observe the oscillatory behavior of correlations, due to the presence of dipole-dipole interaction ${\Omega _{12}}$.   \par
This dynamics can be easily understood from the time evolution of the two equally populated intermediate states, i.e. $\rho_{ss}(0) =\rho_{aa}(0) = 1/2$(see fig. (\ref{fig_2})). It is shown that the  state $|+\rangle$ decays with an enhanced (superradiant) rate while the state $|-\rangle$ decays with the reduced (subradiant) rate. The concurrence exhibits an appreciably long lifetime due to the asymmetry between two cascades, eventually reaching the value of the population of anti-symmetric state, i.e. $C(t) \simeq \rho_{aa}(t)$.
\subsection{Maximally Entangled Symmetric State}
As a third example, let consider the maximally entangled symmetric as the initial atomic state, i.e., 
\begin{equation}
\left| {\psi \left( 0 \right)} \right\rangle  = \frac{1}{{\sqrt 2 }}\left( {\left| {{e_1},{g_2}} \right\rangle  + \left| {{g_1},{e_2}} \right\rangle } \right).
\end{equation}
In this case, the evolved density matrix takes the form
\begin{equation}\label{Matrix2cas}
\rho\left( \tau  \right) = \left( \begin{array}{*{20}{c}}
0&0&0&0\\
0&\alpha\left( \tau  \right)&\alpha\left( \tau  \right)&0\\
0&\alpha\left( \tau  \right)&\alpha\left( \tau  \right)&0\\
0&0&0&\beta\left( \tau  \right)
\end{array} \right),
\end{equation}
with the obtained concurrence and quantum discord
\begin{equation}
C\left(  \tau  \right)=D_T\left( \tau \right) = e^{- \left( 1 + \gamma\right)\tau }.
\end{equation}
On the other hand, to obtain the explicit expression of LQU, we compute first the elements given by Eq. (\ref{w-elements}),
\begin{equation}
w_{11} = w_{22} = \sqrt {e^{ - \left( {1 + \gamma } \right)\tau }w_{33}},
\end{equation}
where $w_{33} = 1 - {e^{ - \left( {1 + \gamma } \right)\tau }}$ and to get it,
one has to treat the cases,  separately, i.e.,  $w_{11}\geq w_{33}$ or $w_{33}\geq w_{11}$ with which
\begin{equation}
\mathcal{U}(\rho\left( \tau  \right)) = 1 - {\rm max}\{ w_{11}, w_{33}\}.
\end{equation}
\begin{figure}[t!]
	\centering
	\begin{minipage}[t]{3in}
		\centering
		\includegraphics[scale=0.55]{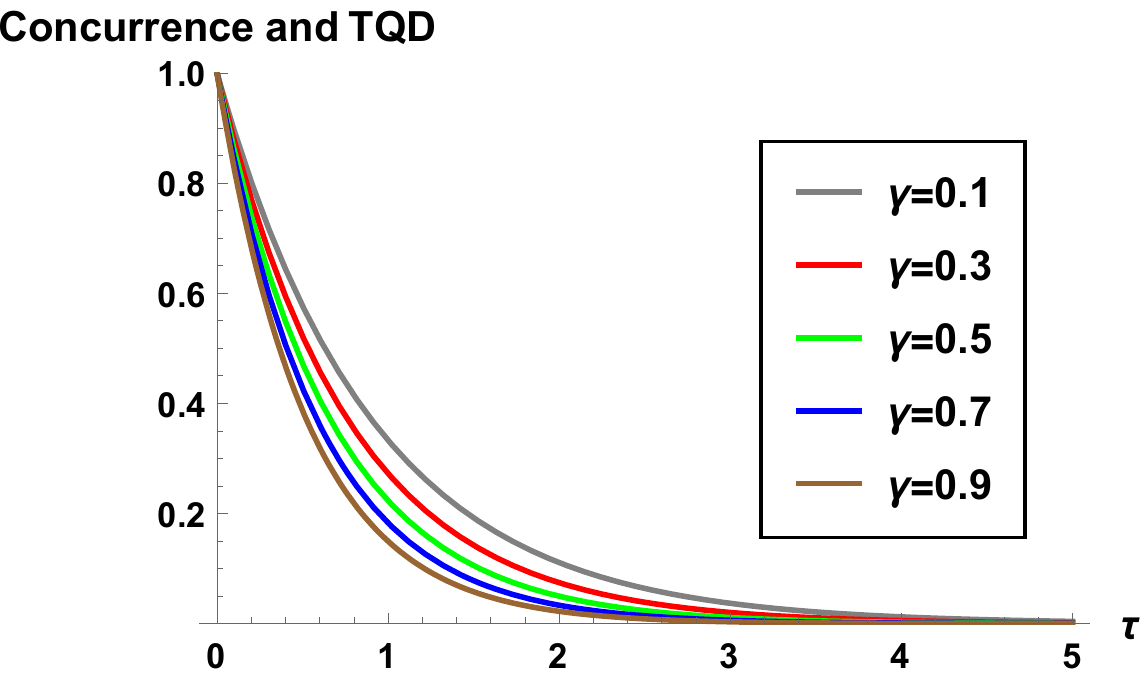}
	\end{minipage}\\
	\caption{Evolutions of concurrence and geometric quantum discord versus the parameter $\tau$ for the different values of $\gamma$.} \label{fig_7}
	\begin{minipage}[t]{3in}
		\centering
		\includegraphics[scale=0.55]{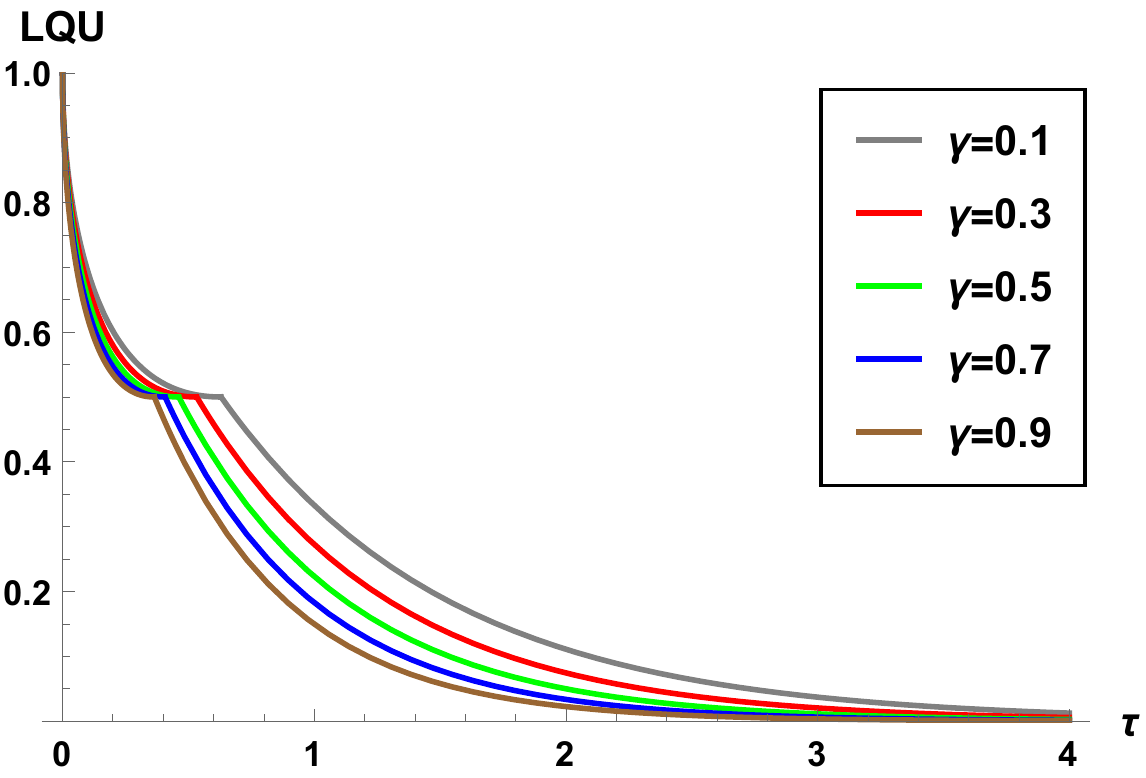}
	\end{minipage}\\
	\caption{Evolution of local quantum uncertainty versus the parameter $\tau$ for the different values of $\gamma$.}
	\label{fig_8}
\end{figure}
The decay of non-classical correlations for the maximally entangled symmetric state is shown in Fig. (\ref{fig_7}) and (\ref{fig_8}). It is observed that the different
measures (concurrence, TQD and LQU) of the correlations have similar evolution behavior and decays rapidly with the enhancement of spontaneous emission parameter $\gamma$.
\subsection{The case of two very close atoms (${r_{12}} \to 0$).}
Here, we will study the behaviour of the concurrence, the trace distance quantum discord and the local quantum uncertainty in the situation when the distance between two atoms tends to zero, i.e., ${r_{12}} \to 0$. Therefore, the collectoive damping ${\Gamma _{12}} \to \Gamma$ and $\Omega _{12} \to \left[ 3\Gamma  \mathord{\left/ \vphantom {3\Gamma  4\xi _{12}^3} \right.
	\kern-\nulldelimiterspace} {4\xi _{12}^3} \right]\left[ 1 - 3\left( {\vec \mu .{\vec r}_{12}} \right)^2 \right]$. In this limit, we shall discuss three cases. we start with the first case in which two atoms in the maximally entangled Bell state for which the concurrence is given by
\begin{equation}
C\left( \tau \right) = \max \left\{ {0,{C_1}\left( \tau  \right),{C_2}\left( \tau  \right)} \right\},
\end{equation}
where
\begin{eqnarray}
C_1\left( \tau  \right) &=& \left( {1 - \tau } \right){e^{ - 2\tau }}, \nonumber \\
C_2\left( \tau  \right) &=& {e^{ - \tau }}\left( {\tau {e^{ - \tau }} - \sqrt {2 - \left( {1 + 2\tau } \right){e^{ - 2\tau }}} } \right).
\end{eqnarray}
Likewise, the trace distance quantum discord and the explicit expression of matrix $W$ are given by
\begin{eqnarray}
D_T\left( \tau \right) &=& {e^{ - \tau }}\left( {1 + \tau {e^{ - \tau }}} \right), \\
w_{11} &=& \frac{{\tau {e^{ - \tau }}A+ \tau {e^{ - 2\tau }}\left( {1 + \sqrt {B} } \right)}}{{\sqrt {\tau \left( {1 + \tau {e^{ - 2\tau }}} \right) + \tau {e^{ - \tau }}\sqrt {B} } }},  \nonumber \\
w_{22} &=& \frac{{\tau {e^{ - \tau }}A+ \tau {e^{ - 2\tau }}\left( {\sqrt {B}  - 1} \right)}}{{\sqrt {\tau \left( {1 + \tau {e^{ - 2\tau }}} \right) + \tau {e^{ - \tau }}\sqrt {B} } }}, \nonumber \\
w_{33} &=& \frac{1}{2}\left( {A + {e^{ - \tau }}\sqrt {B} } \right) \nonumber \\
&+&\frac{{{{\left( {\left( {1 + \tau } \right){e^{ - 2\tau }} - 1} \right)}^2} - {e^{ - 2\tau }}}}{{2\left( {1 - \tau {e^{ - 2\tau }} + {e^{ - \tau }}\sqrt {B} } \right)}}.
\end{eqnarray}
where $A=\left( {1 - \tau {e^{ - 2\tau }}} \right)$ and $B=1 - \left( {1 + 2\tau } \right){e^{ - 2\tau }}$. The second case corresponds to the separable state of singly excited state for which
the concurrence and TQD takes the form
\begin{equation}
C\left( \tau  \right)= D_{T}\left(\tau \right) = \frac{1}{2}\sqrt {{{\left( e^{ - 2\tau } - 1\right)}^2} + e^{- 2\tau}\sin {\left(2\eta \tau \right)^2}}.
\end{equation}
For local quantum uncertainty, the elements of the matrix $W$ are reduced to
\begin{eqnarray}
w_{11} &=& \frac{{\left( {1 - {e^{ - 2\tau }}} \right)\left( {{e^{ - \tau }}\cos \left( {2\eta \tau } \right) - 2{e^{ - 2\tau }} + 2} \right)}}{{4\sqrt {1 - {e^{ - 4\tau }}} }}, \nonumber \\
w_{22} &=& \frac{{\left( {{e^{ - 2\tau }} - 1} \right)\left( {{e^{ - \tau }}\cos \left( {2\eta \tau } \right) + 2{e^{ - 2\tau }} - 2} \right)}}{{4\sqrt {1 - {e^{ - 4\tau }}} }},    \nonumber \\
w_{33} &=& \frac{{{e^{ - 2\tau }}\left( {2\cos \left( {4\eta \tau } \right) - {e^{ - 2\tau }} + 2} \right) + 1}}{{2\left( {{e^{ - 2\tau }} + 1} \right)}},
\end{eqnarray}
Similarly for the third case, in which the initial atomic state is the maximally entangled symmetric state, the concurrence and TQD takes the form
\begin{equation}
	C\left( \tau \right) = {D_T}\left( \tau \right) = {e^{ - 2\tau }},
\end{equation}
and the elements of Eq. (\ref{w-elements}) are reduced to
\begin{equation}
w_{11} = w_{22} = {e^{ - \tau }}\sqrt {w_{33}},
\end{equation}
where $w_{33} = 1 - e^{ - 2\tau }$.
\begin{figure}[t!]
	\centering
	\begin{minipage}[t]{3in}
		\centering
		\includegraphics[scale=0.4]{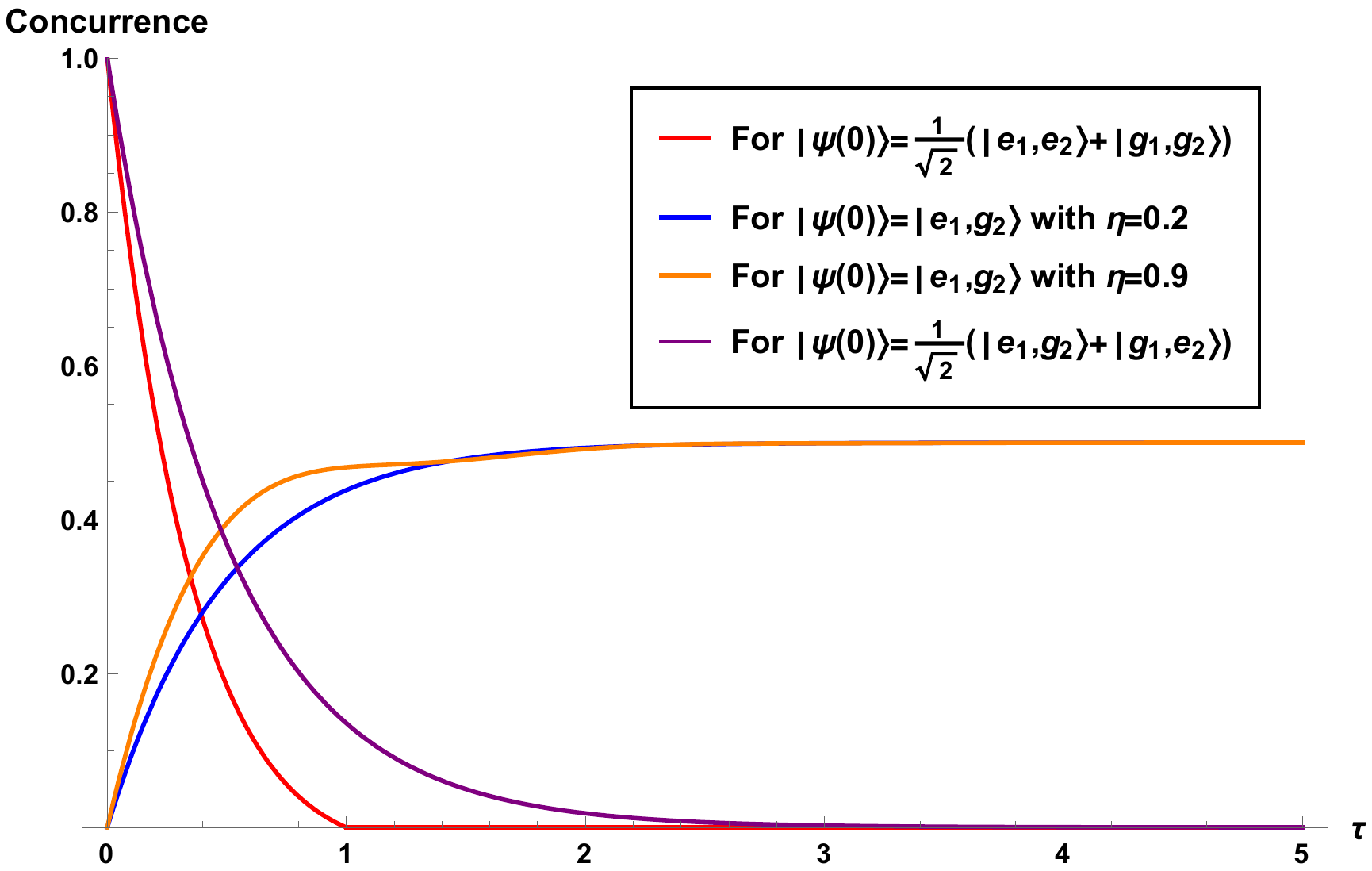}
	\end{minipage}\\
	\caption{Evolutions of concurrence for the different cases versus the parameter $\tau$ with $\gamma \longrightarrow 1$ }\label{fig_9}
	\begin{minipage}[t]{3in}
		\centering
		\includegraphics[scale=0.4]{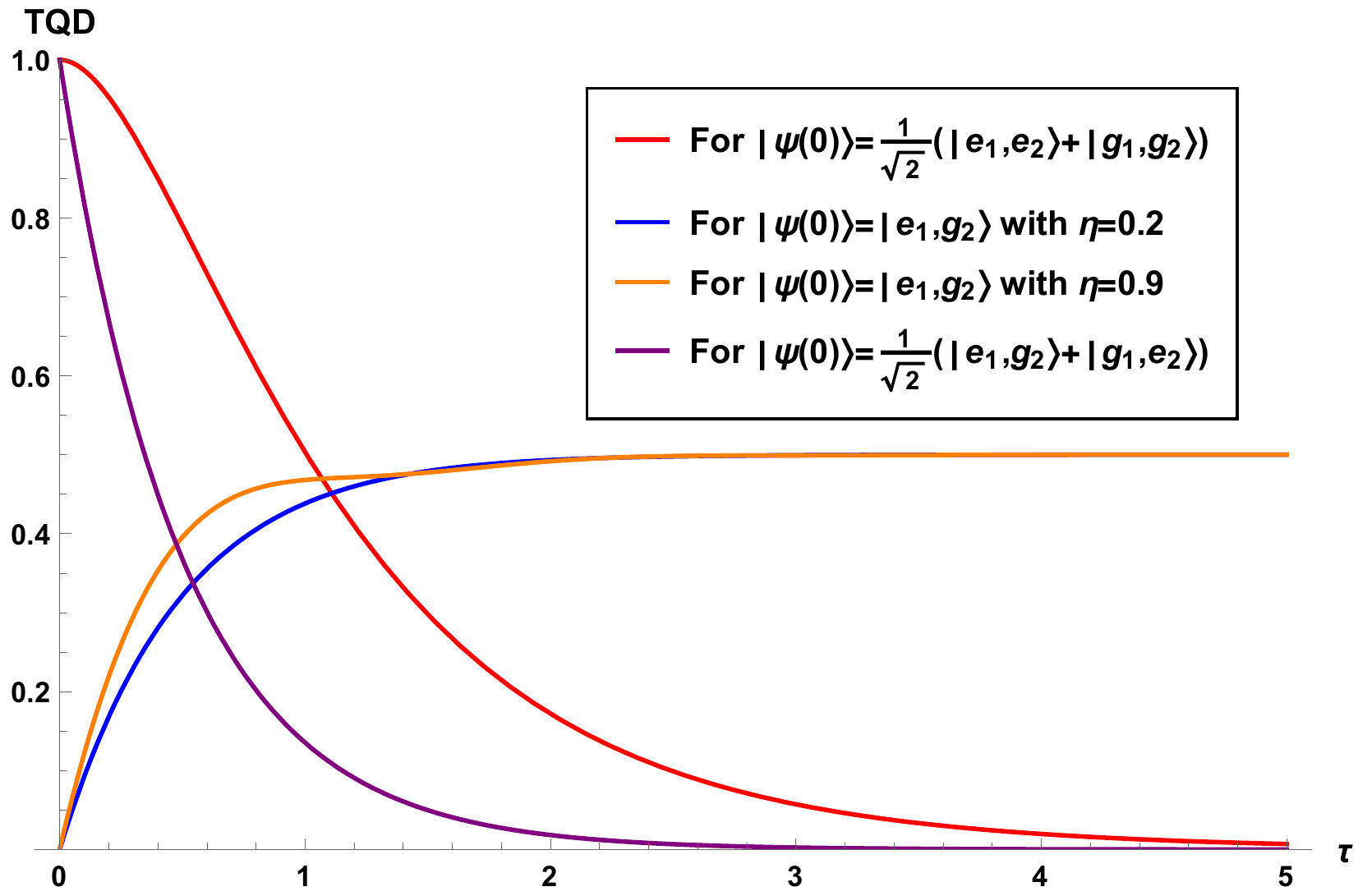}
	\end{minipage}\\
	\caption{Evolutions of the trace quantum
		discord for the different cases versus the parameter $\tau$ with $\gamma \longrightarrow 1$ }\label{fig_10}
	\begin{minipage}[t]{3in}
		\centering
		\includegraphics[scale=0.4]{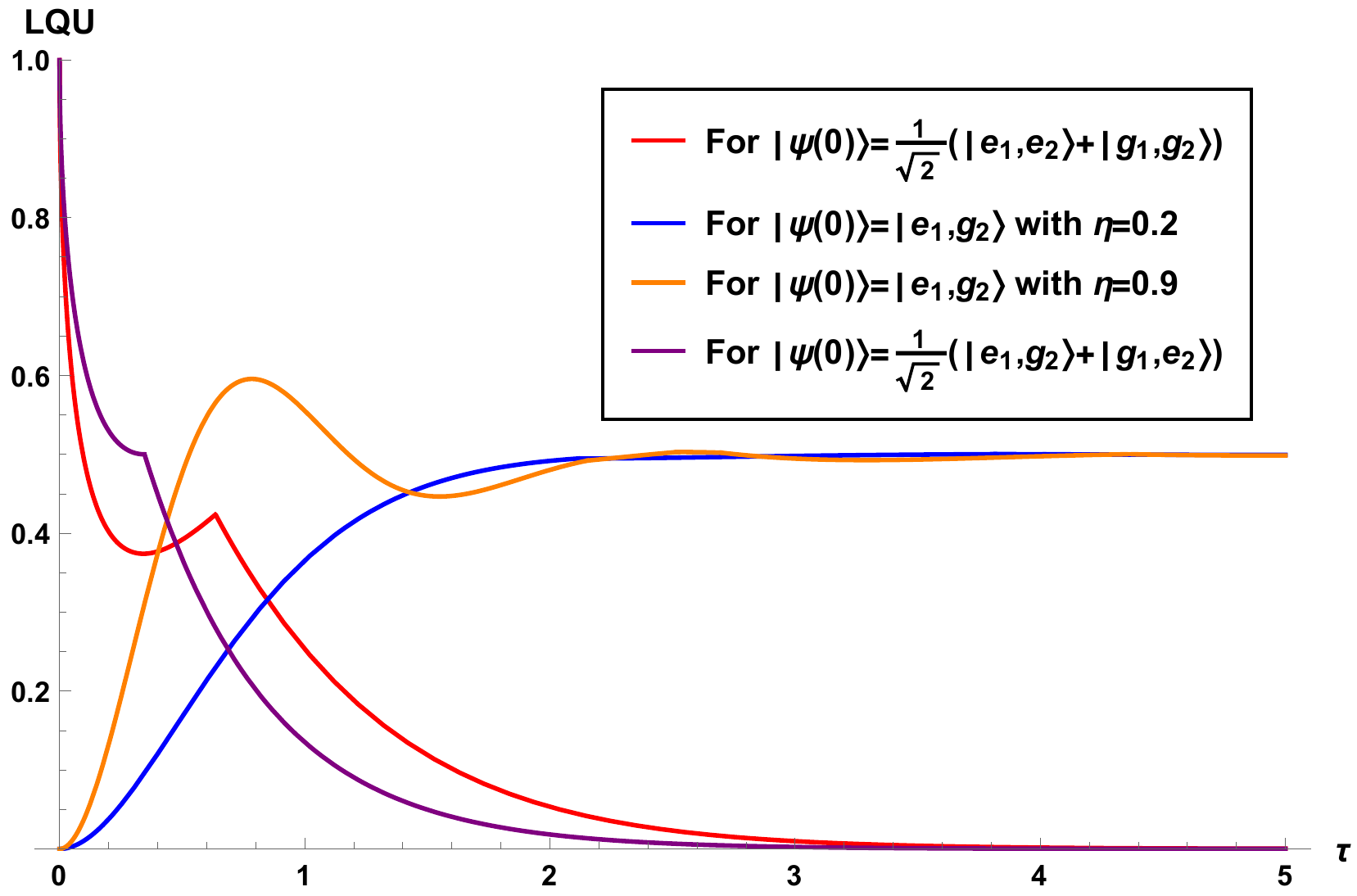}
	\end{minipage}\\
	\caption{Evolutions of the local quantum uncertainty for the different cases versus the parameter $\tau$ with $\gamma \longrightarrow 1$ } \label{fig_11}
\end{figure}
To get more insight on the effect of dipole-dipole interaction and collective damping, let assume that the two atoms are approaching one another, i.e., ${r_{12}} \to 0$. We investigate the evolution of quantum correlations in Fig.'s (\ref{fig_9})-(\ref{fig_11}). Indeed, we observe a sudden change in quantum correlations that depend on the inter-atomic distance and initial excitations. Our result shows that the quantum correlations can be maintained over a longer time but for small distances between the two atoms.
\section{Conclusion}
To conclude, we have investigated the dynamics of non-classical correlations in two-two level atoms interacting with a quantized electromagnetic field (assumed to be in the vacuum state). The peculiar behavior of quantum correlations are examined for different initial Dicke states. It depends on the collective damping and qubit-qubit interaction. When the two atoms are initially in an entangled state of double or zero excitation, the evolution of different measures behave differently and decays exponentially. Meanwhile, the entanglement is generated, spontaneously with  oscillatory behavior, by assuming initially the superposition of intermediate states for which all the correlations display their dependence on both collective damping and qubit-qubit interaction. We also have discussed the evolution of these
correlations for the atoms placed very close, i.e., $r_{12} \rightarrow 0$. In this case, all measures shows the long time non-classical correlations which behave identically.
Therefore, the present investigation suggests that local quantum uncertainity and trace quantum discord are good measures of the quantum correlation.
\appendix 
\section{LQU for States of $X$ type}
In this appendix, we present a simple method to calculate the local quantum uncertainty for $X$-type states. First, the eigenvalues corresponding to Eq. (\ref{X}) are,
\begin{eqnarray}
{\lambda _1} = \frac{1}{2}{t_1} + \frac{1}{2}\sqrt {{t_1}^2 - 4{d_1}} , \qquad
{\lambda _2} = \frac{1}{2}{t_2} + \frac{1}{2}\sqrt {{t_2}^2 - 4{d_2}}  \nonumber \\
{\lambda _3} = \frac{1}{2}{t_2} - \frac{1}{2}\sqrt {{t_2}^2 - 4{d_2}} , \qquad
{\lambda _4} = \frac{1}{2}{t_1} - \frac{1}{2}\sqrt {{t_1}^2 - 4{d_1}}  \nonumber \\
\end{eqnarray}
where
\begin{equation}
\left\{ \begin{array}{l}
{t_1} = {\rho _{11}} + {\rho _{44}}\\
{d_1} = {\rho _{11}}{\rho _{44}} - {\rho _{14}}{\rho _{41}}\\
{t_2} = {\rho _{22}} + {\rho _{33}}\\
{d_2} = {\rho _{22}}{\rho _{33}} - {\rho _{32}}{\rho _{23}}
\end{array} \right.
\end{equation}
The square root of Eq. (\ref{X}) in terms of the computational basis can be written as follows,
\begin{equation}
{\scriptsize \sqrt \rho   = \left( {\begin{array}{*{20}{c}}
		{\frac{{{\rho _{11}} + \sqrt {{d_1}} }}{{\sqrt {{t_1} + 2\sqrt {{d_1}} } }}}&0&0&{\frac{{{\rho _{14}}}}{{\sqrt {{t_1} + 2\sqrt {{d_1}} } }}}\\
		0&{\frac{{{\rho _{22}} + \sqrt {{d_2}} }}{{\sqrt {{t_2} + 2\sqrt {{d_2}} } }}}&{\frac{{{\rho _{23}}}}{{\sqrt {{t_2} + 2\sqrt {{d_2}} } }}}&0\\
		0&{\frac{{{\rho _{32}}}}{{\sqrt {{t_2} + 2\sqrt {{d_2}} } }}}&{\frac{{{\rho _{33}} + \sqrt {{d_2}} }}{{\sqrt {{t_2} + 2\sqrt {{d_2}} } }}}&0\\
		{\frac{{{\rho _{41}}}}{{\sqrt {{t_1} + 2\sqrt {{d_1}} } }}}&0&0&{\frac{{{\rho _{44}} + \sqrt {{d_1}} }}{{\sqrt {{t_1} + 2\sqrt {{d_1}} } }}}
		\end{array}} \right).} \label{sqrt density matrix}
\end{equation}
with the eigenvalues
\begin{eqnarray}
\sqrt {{\lambda _1}} &=& \frac{1}{2}\sqrt {{t_1} + 2\sqrt {{d_1}} }  + \frac{1}{2}\sqrt {{t_1} - 2\sqrt {{d_1}} }, \nonumber \\
\sqrt {{\lambda _2}} &=& \frac{1}{2}\sqrt {{t_2} + 2\sqrt {{d_2}} }  + \frac{1}{2}\sqrt {{t_2} - 2\sqrt {{d_2}} }, \nonumber \\
\sqrt {{\lambda _3}}  &=& \frac{1}{2}\sqrt {{t_2} + 2\sqrt {{d_2}} }  - \frac{1}{2}\sqrt {{t_2} - 2\sqrt {{d_2}} }, \nonumber \\
\sqrt {{\lambda _4}}  &=& \frac{1}{2}\sqrt {{t_1} + 2\sqrt {{d_1}} }  - \frac{1}{2}\sqrt {{t_1} - 2\sqrt {{d_1}} } .
\end{eqnarray}
The density matrix operator $\sqrt {{\rho}}$ of Eq. (\ref{sqrt density matrix}) can be described in Fano-Bloch representation as
\begin{equation}
\sqrt \rho   = \frac{1}{4}\sum\limits_{\chi ,\delta } {{R_{\chi \delta }}} {\sigma _\chi } \otimes {\sigma _\delta },
\end{equation}
where the parameters of the correlation matrix ${R_{\chi \delta }} = tr\left( {\sqrt \rho  {\sigma _\chi } \otimes {\sigma _\delta }} \right)$ with $\chi ,\delta  = 0,1,2,3$ are,
\begin{eqnarray}
{R_{00}} &=& \sqrt {{t_1} + 2\sqrt {{d_1}} }  + \sqrt {{t_2} + 2\sqrt {{d_2}} } \nonumber \\
{R_{03}} &=& \frac{1}{2}\frac{{{T_{30}} + {T_{03}}}}{{\sqrt {{t_1} + 2\sqrt {{d_1}} } }} - \frac{1}{2}\frac{{{T_{30}} - {T_{03}}}}{{\sqrt {{t_2} + 2\sqrt {{d_2}} } }}   \nonumber \\
{R_{30}} &=& \frac{1}{2}\frac{{{T_{30}} + {T_{03}}}}{{\sqrt {{t_1} + 2\sqrt {{d_1}} } }} + \frac{1}{2}\frac{{{T_{30}} - {T_{03}}}}{{\sqrt {{t_2} + 2\sqrt {{d_2}} } }}  \nonumber \\
{R_{11}} &=& \frac{1}{2}\frac{{{T_{11}} + {T_{22}}}}{{\sqrt {{t_2} + 2\sqrt {{d_2}} } }} + \frac{1}{2}\frac{{{T_{11}} - {T_{22}}}}{{\sqrt {{t_1} + 2\sqrt {{d_1}} } }} \nonumber \\
{R_{12}} &=& \frac{1}{2}\frac{{{T_{12}} - {T_{21}}}}{{\sqrt {{t_2} + 2\sqrt {{d_2}} } }} + \frac{1}{2}\frac{{{T_{12}} + {T_{21}}}}{{\sqrt {{t_1} + 2\sqrt {{d_1}} } }}  \nonumber \\
{R_{21}} &=& \frac{1}{2}\frac{{{T_{12}} + {T_{21}}}}{{\sqrt {{t_1} + 2\sqrt {{d_1}} } }} - \frac{1}{2}\frac{{{T_{12}} - {T_{21}}}}{{\sqrt {{t_2} + 2\sqrt {{d_2}} } }} \nonumber \\
{R_{22}} &=& \frac{1}{2}\frac{{{T_{11}} + {T_{22}}}}{{\sqrt {{t_2} + 2\sqrt {{d_2}} } }} - \frac{1}{2}\frac{{{T_{11}} - {T_{22}}}}{{\sqrt {{t_1} + 2\sqrt {{d_1}} } }}  \nonumber \\
{R_{33}} &=& \sqrt {{t_1} + 2\sqrt {{d_1}} }  - \sqrt {{t_2} + 2\sqrt {{d_2}} }.
\end{eqnarray}
Reposting the expression of $\sqrt {{\rho}}$ (\ref{sqrt density matrix}) in Eq. (\ref{w-elements}), one gets, after some algebra, the matrix elements $w_{ij}$ given by Eq. (\ref{w11}).
\section*{Acknowledgements}
MIS acknowledges the support from the DP-PMI programme and Funda\c{c}\~{a}o para a Ci\^{e}ncia e a Tecnologia (Portugal), namely through the scholarship number SFRH/PD/BD/113650/2015.


\begin{thebibliography}{99}
	\bibitem{Einstein1935}A. Einstein, B. Podolsky and N. Rosen, Phys. Rev. \textbf{47} (1935) 777.
	\bibitem{Bell1964}Bell, J. S. (1966). Physics 1, \textbf{195} (1964).
	\bibitem{Hill1997}Hill, S., Wootters, W. K. (1997). Phys Rev lette,\textbf{78}(26), 5022.
	\bibitem{Ekert1991} A. K. Ekert, Phys.Rev. Lett. {\bf 67}, 661 (1991). 
	\bibitem{Bennett1992}C. H. Bennett and S. J. Wiesner, Phys. Rev. Lett. {\bf 69}, 2881 (1992). 
	\bibitem{Bennett1993} C. H. Bennett, G. Brassard, C. Cepeau, R. Jozsa, A. Peres and W. K.   Wootters, Phys. Rev. Lett. {\bf 70}, 1895 (1993). 
	
	\bibitem{Olliver2001} H. Olliver and W. H. Zurek, Phys. Rev. Lett. \textbf{88}, 017901 (2001).
	\bibitem{Modi2012} K. Modi, A. Brodutch, H. Cable, T. Paterek, and V. Vedral, Rev. Mod. Phys. \textbf{84},1655 (2012).
	\bibitem{Céleri2011}L. C. Céleri, J. Maziero, and R. M. Serra, Int. J. Quantum Inform. \textbf{09}, 1837 (2011).
	
	\bibitem{Luo2008} S. Luo, Phys. Rev. A \textbf{77}, 042303 (2008).
	\bibitem{Ali2010} M. Ali, A. R. P. Rau, and G. Alber, Phys. Rev. A \textbf{81}, 042105 (2010).
	\bibitem{Tanaś2013} R. Tanaś, Phys. Scr. T \textbf{153}, 014059 (2013).
	
	
	\bibitem{Chuan2012}T. K. Chuan, J. Maillard, K. Modi, T. Paterek, M. Paternostro, and M. Piani, Phys. Rev. Lett. \textbf{109}, 070501 (2012).
	\bibitem{Streltsov2012} A. Streltsov, H. Kampermann, and D. Bruß, Phys. Rev. Lett. \textbf{108}, 250501 (2012).
	
	\bibitem{Madhok2011} V. Madhok and A. Datta, Phys Rev A  \textbf{83}, 032323 (2011).	
	\bibitem{Cavalcanti2011} D. Cavalcanti, L. Aolita, S. Boixo, K. Modi, M. Piani, and A. Winter, Phys Rev A \textbf{83}, 0323248 (2011).
	
	\bibitem{Dicke1954} R.H. Dicke, Phys. Rev. {\bf 93}, 99 (1954).
	\bibitem{Tavis1968} M.  Tavis  and  F.  W.  Cummings,  Phys.  Rev. {\bf 170}, 379 (1968).

	\bibitem{Lukin2001} Mikhail Lukin, Michael Fleischhauer and Atac Imamoğlu "\textit{Quantum information processing based on cavity QED with mesoscopic systems}", p. 193, Springer Berlin Heidelberg, (2001).
	\bibitem{Steane1997}A. M. Steane, Appl.Phys. B {\bf 64}, 623 (1997).
	\bibitem{Peter2017}Peter Michler, \textit{Quantum dots for quantum information technologies}, Springer international publishing, (2017).
	
	\bibitem{Braun2002}D. Braun,, Phys. Rev. Lett. {\bf 89}, 277901 (2002).
	\bibitem{Yu2004}T. Yu and J. H. Eberly, Phys. Rev. Lett. {\bf 93}, 140404 (2004).
	\bibitem{Muzzamal2013} M. I. Shaukat, A. Shaheen and A.H. Toor, J. of Mod. Opt. {\bf 60}, 21 (2013).
	\bibitem{Ficek2006} Z. Ficek and R. Tanas, Phys. Rev. A {\bf 74}, 024304 (2006).
	\bibitem{Ficek2008} Z. Ficek and R. Tanas, Phys. Rev. A {\bf 77}, 054301 (2008).
	\bibitem{Verstraete2009} F. Verstraete, M. M. Wolf and I. Cirac, Nature Phys. {\bf 5}, 633 (2009).
	
	\bibitem{muzzamal2018} M. I. Shaukat, E. V. Castro and H. Ter\c{c}as, arXiv:1801.08894 (2018).
	\bibitem{He2017}  Y. He and M. Jiang, Opt. Comm. {\bf 382}, 580 (2017).
	\bibitem{Muzzamal2018} M. I. Shaukat, E. V. Castro and H. Ter\c{c}as, arXiv:1801.08169 (2018).
	\bibitem{Muzzamal2017} M. I. Shaukat, E. V. Castro and H. Ter\c{c}as, Phys. Rev. A \textbf{95}, 053618 (2017).
	\bibitem{Khulud2011} K. Almutairi, R. Tana and Z. Ficek, Phys. Rev. A \textbf{84}, 013831 (2011).
	
	
	
	\bibitem{Wang2010} B. Wang, Z. Y. Xu, Z. Q. Chen, and M. Feng, Phys. Rev. A \textbf{81}, 014101 (2010).
	\bibitem{Maziero2009} J. Maziero, L. C. Céleri, R. M. Serra, V. Vedral, Phys. Rev. A \textbf{80}, 044102 (2009).
	\bibitem{Bellomo2012} B. Bellomo, G. L. Giorgi, F. Galve, R. Lo Franco, G. Compagno, R. Zambrini, Phys. Rev. A \textbf{85}, 032104 (2012).
	\bibitem{Pinto2013} J. P. G.. Pinto, G. Karpat, and F. F. Fanchini, Phys. Rev. A \textbf{88}, 034304 (2013).
	\bibitem{Aaronson2013} B. Aaronson, R. Lo Franco, and G. Adesso, Phys. Rev. A \textbf{88}, 012120 (2013).
	\bibitem{Hu2014} M.-L. Hu, D.-P. Tian, Ann. Phys. (NY) \textbf{343}, 132 (2014).
	
	\bibitem{Knill1998} E. Knill and R. Laflamme, Phys. Rev. Lett. \textbf{81}, 5672 (1998).	
	
	\bibitem{Lu2011} X. M. Lu, J. Ma, Z. Xi and X. Wang, Phys. Rev. A \textbf{83}, 012327 (2011)
	\bibitem{Chen2011} Q. Chen, C. Zhang, S. Yu, X. X. Yi and C. H. Oh, Phys. Rev. A \textbf{84}, 042313 (2011).
	
	\bibitem{Wootters1998} W. K. Wootters, Phys. Rev. Lett. {\bf 80}, 2245 (1998).
	\bibitem{Bennett1996}C. H. Bennett, H. J. Bernstein, S. Popescu and B. Schumacher, Phys. Rev. A \textbf{53} (1996) 2046.
	\bibitem{Popescu1997}S. Popescu and D. Rohrlich, Phys. Rev. A \textbf{56} (1997) R3319.
	\bibitem{Wootters2001}Wootters, W. K. (2001). Quant. Inf. Comp \textbf{1}(2001), 27-44.
	\bibitem{Yu2009}T. Yu and J. H. Eberly, Science \textbf{323} (2009) 598.
	\bibitem{Bose2000}S. Bose and V. Vedral, Phys. Rev. A \textbf{61} (2000) 040101.
	\bibitem{Brassard1996}C. H. Bennett, G. Brassard, S. Popescu, B. Schumacher, J. Smolin and W. K. Wootters, Phys. Rev. Lett. \textbf{76} (1996) 722.
	\bibitem{Peres1996}A. Peres, Phys. Rev. Lett. \textbf{77} (1996) 1413.
	\bibitem{Vidal2002}G. Vidal and R. F. Werner, Phys. Rev. A \textbf{65} (2002) 032314.
	\bibitem{Ollivier2001}H. Ollivier and W. H. Zurek, Phys. Rev. Lett. \textbf{88} (2001) 017901.
	\bibitem{Henderson2001}L. Henderson and V. Vedral, J. Phys. A \textbf{34} (2001) 6899.
	\bibitem{Dakic2010} B. Daki\'{c}, V. Vedral, and \v{C}. Brukner, Phys. Rev. Lett. \textbf{105} (2010) 190502.
	\bibitem{Bellomo1} B. Bellomo,  R. Lo Franco and G. Compagno, Phys. Rev. A \textbf{86} (2012) 012312.
	
	\bibitem{Bellomo2}B. Bellomo, G.L. Giorgi, F. Galve, R. Lo Franco, G. Compagno and R. Zambrini, Phys. Rev. A (2012) 032104.
	
	\bibitem{DaoudPLA}M. Daoud and R. Ahl Laamara, Phys. Lett. A (2012) 2361.
	
	\bibitem{DaoudIJQI}M. Daoud and R. Ahl Laamara, Int. J.  Quantum Inf.\textbf{10} (2012) 1250060.
	\bibitem{piani} M. Piani, Phys. Rev. A {\bf 86} (2012) 034101.
	\bibitem{paula2013} F. M. Paula, T. R. de Oliveira, and M. S. Sarandy, Phys.
	Rev. A {\bf 87} (2013) 064101; F.M. Paula, J.D. Montealegre, A. Saguia, T.R. de Oliveira and M.S. Sarandy, EPL, \textbf{103} (2013) 50008.
	\bibitem{Bromley2014} T.R.  Bromley, M. Cianciaruso, R. Lo Franco, G. Adesso, J. Phys. A: Math. Theor. {\bf 47} (2014) 405302
	\bibitem{Ciccarello2014} F. Ciccarello, T. Tufarelli and V. Giovannetti, New J. Phys. \textbf{ 16} (2014) 013038.
	\bibitem{Girolami2013}D. Girolami, T. Tufarelli, and G. Adesso, Phys. Rev. Lett. \textbf{110} (2013) 240402.
	\bibitem{Wigner1963}E. P. Wigner and M. M. Yanasse, Proc. Nat. Acad. Sci. USA \textbf{49} (1963) 910.
	\bibitem{Luo2003}S. Luo, Phys. Rev. Lett. 91 (2003) 180403.
	\bibitem{Agarwal1974} G. Agarwal, Quantum Statistical Theories of Spontaneous Emission and their Relation to Other Approaches, vol. 70 of Springer Tracts in Modern Physics (Springer-Verlag, Berlin, 1974).
	
	\bibitem{Tanas2004} A. Slaoui,  M. Daoud, and R.A. Laamara. Quantum Information Processing, \textbf{17} (2018) 178.
	
	\bibitem{Ficek1987} Z.  Ficek, R. Tana${\rm \acute{s}}$ and  S. Kielich, Physica A {\bf 146} (1987) 452.
	
	\bibitem{Auyuanet2010} A. Auyuanet and  L. Davidovich, Phys. Rev.  A {\bf 82} (2010) 032112.
	\bibitem{Lehmberg1970}R.H. Lehmberg, Phys. Rev. A \textbf{2}, 883; \textbf{2}, 889 (1970).
	
	\bibitem{Belavkin1969}Belavkin, A. A., Zeldovich, B. Y., Perelomov, A. M., Popov, V. S. (1969). Sov. Phys. JETP, \textbf{56}, 264-274.
	
	\bibitem{Agarwal1970}Agarwal, G. S.(1970). Phys Rev A, \textbf{2(5)}, 2038. ISO 690.	
	

\end{thebibliography}
\end{document}